\documentclass[11pt]{article}
\usepackage{times,amsmath,amsthm,amssymb,graphicx,hyperref,float,xcolor}
\usepackage{fancyhdr}     
\usepackage{setspace}    
\usepackage{booktabs}    
\usepackage{longtable}   
\usepackage{multirow}
\usepackage{textcomp}    
\usepackage{bm}         
\usepackage{pdfpages}    
\usepackage{rotating}
\usepackage[percent]{overpic} 

\newtheorem{nt}{Note}

\newtheorem{xmpl}{Example}

\def \ds{\displaystyle}

\newcommand{\bbeta}{\bm{\beta}}
\newcommand{\bPe}{\bm{Pe}}
\newcommand{\bchi}{\bm{\chi}}
\newcommand{\R}{\mathbb R}
\newcommand{\N}{\mathbb N}

\begin{document}
\global\def\refname{{\normalsize \it References:}}
\baselineskip 12.5pt
%
%
%
\title{\LARGE \bf Theoretical Analysis of a Two-Dimensional Bilayer Convection-Diffusion-Reaction-Source Problem}

\date{}

\author{\hspace*{-10pt}
\begin{minipage}[t]{2.3in} \normalsize \baselineskip 12.5pt
\centerline{GUILLERMO FEDERICO UMBRICHT}
\centerline{LIDTUA, Facultad de Ingenier\'ia, Universidad Austral}
\centerline{Mariano Acosta 1611, Pilar, Buenos Aires, ARGENTINA}
\vspace{0.5cm}
\centerline{DIANA RUBIO}
\centerline{ITECA (UNSAM-CONICET), CEDEMA, ECyT, Universidad
Nacional de General San Mart\'in}
\centerline{25 de mayo y Francia, San Mart\'in, Buenos Aires, ARGENTINA}
\vspace{0.5cm}
\centerline{DOMINGO ALBERTO TARZIA}
\centerline{Departamento de Matem\'atica, Facultad de Ciencias Empresariales, Universidad Austral}
\centerline{Paraguay 1950, Rosario, Santa Fe, ARGENTINA}
\centerline{Consejo Nacional de Investigaciones Cient{\'i}ficas y T\'ecnicas (CONICET)}
\centerline{Godoy Cruz 2290, CABA, ARGENTINA}
\end{minipage}
%
%
\\ \\ \hspace*{-10pt}
\begin{minipage}[b]{6.9in} \normalsize
\baselineskip 12.5pt {\it Abstract:}
This work investigates the two-dimensional thermal behavior of a bilayer medium subject to both internal and external heat sources. The model incorporates diffusion, advection, and temperature-dependent volumetric heat generation or absorption in each layer, as well as general convective conditions on the external boundaries. The influence of interfacial thermal resistance between the two materials is also considered. An analytical solution is developed using Fourier-based techniques, and a stable and convergent finite difference method is proposed to analyze particular scenarios. The theoretical results are validated against known solutions and numerical simulations, demonstrating consistency with the expected physical behavior. The findings contribute to a deeper understanding of heat transfer phenomena in layered systems and offer potential insights for optimizing thermal performance in engineering applications involving composite materials.
\\ [4mm] {\it Key--Words:}
Heat transfer, Multilayer materials, Two-dimensional analysis, Interfacial thermal resistance.\\
%
\end{minipage}
\vspace{-10pt}}

\maketitle

\thispagestyle{empty} \pagestyle{empty}
%
%

\section*{Nomenclature}

\begin{longtable}{p{11mm} c p{120mm} }
\multicolumn{3}{l}{\textbf{Subscripts and Superscripts}}\\
\\
$0$ & --- & initial value\\
$\infty$ & --- & stationary state\\
$A$ & --- & relative to the left boundary\\
$B$ & --- & relative to the right boundary\\
$H$ & --- & homogeneous system\\
$i$ & --- & spatial grid position (numerical method) \\
$j$ & --- & spatial grid position (numerical method) \\
$k$ & --- & time grid position (numerical method) \\
$m \, (1,2) $ & --- & layer number\\
$n$ & --- & eigenvalue number\\
$p$ & --- & eigenvalue number\\
$x$ & --- & indicates in the $x$ direction \\
$y$ & --- & indicates in the $y$ direction \\

\\
\multicolumn{3}{l}{\textbf{Capital Letters}}\\
\\

$A$ & --- & auxiliary dimensionless parameter\\
$B$ & --- & auxiliary dimensionless parameter\\
$\bar{A}$ & --- & auxiliary temporal function\\
$\bar{B}$ & --- & auxiliary temporal function\\
$Bi$ & --- & Biot number \\
$Bi^*$ & --- & auxiliary dimensionless parameter \\
$\bar{Bi}$ & --- & auxiliary dimensionless parameter\\
$C$ & --- & the specific heat at constant pressure $\textbf{[J(kg{$^{\circ}$}C)$^{-1}$]}$\\
$D$ & --- & differential operator. \\
$\bar{D}$ & --- & dimensionless differential operator\\
$K$ & --- & auxiliary dimensionless parameter (Initial condition)\\
$Pe$ & --- & Péclet number \\
$P$     & --- & auxiliary function (numerical method) $\textbf{[{$^{\circ}$}C]}$\\
$\mathcal{P}$     & --- & partition (numerical method)\\
$R$ & --- & related to the thermal resistance at the interface $\textbf{[m]}$\\
$\bar{R}$ & --- & dimensionless $R$\\
$S$    & --- & auxiliary dimensionless heat source\\
$T$    & --- & temperature field relative to ambient $\textbf{[{$^{\circ}$}C]}$\\
$T_r$ & --- & reference temperature $\textbf{[{$^{\circ}$}C]}$\\

\\
\multicolumn{3}{l}{\textbf{Lowercase Letters} }\\
\\
$f$ & --- & dimensionless auxiliary spatial function\\
$g$ & --- & dimensionless auxiliary temporal function\\
$h$     & --- & convection heat transfer coefficient $\textbf{[Wm{$^{-2}$}({$^{\circ}$}C)$^{-1}$]}$\\
$q$ & --- & auxiliary parameter\\
$s$  & --- & heat source $\textbf{[{$^{\circ}$}C\,s$^{-1}$]}$\\
$\bar{s}$ & --- & dimensionless auxiliary heat source\\
$\widehat{s}$ & --- & dimensionless heat source\\
$t$     & --- & temporary variable $\textbf{[s]}$\\
$u$ & --- & dimensionless auxiliary spatial function\\
$w$ & --- & body length in the $y$ direction $\textbf{[m]}$\\
$\bar{w}$ & --- & dimensionless body length in the $y$ direction \\
$x$     & --- & spatial variable in the $x$ direction $\textbf{[m]}$\\
$\bar{x}$     & --- & dimensionless spatial variable in the $x$ direction\\
$x_1$ & --- & interface location $\textbf{[m]}$\\
$\bar{x}_1$ & --- & dimensionless interface location\\
$x_2$ & --- & body length in the $x$ direction $\textbf{[m]}$\\
$y$     & --- & spatial variable in the $y$ direction $\textbf{[m]}$\\
$\bar{y}$     & --- & dimensionless spatial variable in the $y$ direction\\
$z$     & --- & auxiliary parameter (numerical method)\\

\\
\multicolumn{3}{l}{\textbf{Greek Letters}}\\
\\

$\alpha$ & --- & thermal diffusivity coefficient $\textbf{[m{$^{2}$}s$^{-1}$]}$\\
$\bar{\alpha}$ & --- & dimensionless thermal diffusivity coefficient\\
$\bbeta$ & --- & fluid velocity $\textbf{[m\,s$^{-1}$]}$\\
$\bchi$ & --- & auxiliary dimensionless parameter\\
$\delta$ & --- & auxiliary dimensionless parameter\\
$\nabla$     & --- & Nabla operator\\
$\Delta$     & --- & Laplacian operator\\
$\Delta t$     & --- & time discretization step (numerical method) $\textbf{[s]}$\\
$\Delta x$     & --- & spatial discretization step (numerical method) $\textbf{[m]}$\\
$\Delta y$     & --- & spatial discretization step (numerical method) $\textbf{[m]}$\\
$\epsilon$     & --- & dimensionless spatial eigenvalue\\
$\eta$ & --- & auxiliary dimensionless parameter\\
$\gamma$ & --- & auxiliary dimensionless parameter\\
$\iota$     & --- & auxiliary parameter (numerical method)\\
$\kappa$ & --- & thermal conductivity coefficient $\textbf{[W(m{$^{\circ}$}C)$^{-1}$]}$\\
$\bar{\kappa}$ & --- & dimensionless thermal conductivity coefficient\\
$\lambda$     & --- & dimensionless temporal eigenvalue\\
$\Lambda$ & --- & auxiliary parameter (numerical method) $\textbf{[W(m{$^{\circ}$}C)$^{-1}$]}$\\
$\mu$     & --- & auxiliary dimensionless parameter\\
$\nu$ & --- & generation/consumption coefficient $\textbf{[s$^{-1}$]}$\\
$\bar{\nu}$ & --- & dimensionless generation/consumption coefficient\\
$\omega$     & --- & dimensionless spatial eigenvalue\\
$\Omega$     & --- & problem definition domain\\
$\bar{\Omega}$     & --- & dimensionless problem definition domain\\
$\phi$ & --- & auxiliary dimensionless parameter\\
$\Pi$     & --- & auxiliary parameter (numerical method) $\textbf{[m$^{-1}$]}$\\
$\psi$     & --- & auxiliary dimensionless parameter\\
$\rho$ & --- & density $\textbf{[kg m$^{-3}$]}$\\
$\sigma$ & --- & auxiliary dimensionless parameter\\
$\tau$    & --- & dimensionless temporary variable\\
$\theta$ & --- & dimensionless temperature\\
$\Theta$ & --- & dimensionless auxiliary temperature function\\
$\upsilon$     & --- & auxiliary parameter (numerical method)\\
$\Upsilon$     & --- & auxiliary parameter (numerical method)\\
$\varphi$    & --- & auxiliary dimensionless parameter\\
$\xi$ & --- & auxiliary dimensionless parameter\\
$\zeta$     & --- & auxiliary parameter (numerical method)\\

\end{longtable}

\section{Introduction}
\label{Introduccion} \vspace{-4pt}

The study of thermal and mass transport in multilayered systems has attracted significant attention in recent years, owing to its wide-ranging relevance across engineering, industrial, and scientific applications \cite{Yuan22,Carson22,Zhou21,Zhou21b,Cengel07,Yavaraj23}. These layered configurations appear in diverse contexts such as composite materials, porous media, electronic devices, biological tissues, and energy systems \cite{Hickson09, Hahn12,Kakac85, Cavalcante08, Becker13, Pasupuleti11, Choobineh15}. Research contributions span from textile engineering \cite{Caunce08} and environmental remediation \cite{French81,Liu98,Liu08, Ferragut13}, to biomedical applications like transdermal drug delivery and tumor modeling \cite{Mitragotri11,McGinty11,McGinty14,Jain22,Mantzavinos16}, and thermal management in high-performance systems \cite{Bandhauer11, Shah16, Shah16b, Krishnan21, Subramanian04}.

A wide variety of mathematical models and solution techniques have been developed to describe such problems. Analytical approaches such as the separation of variables, recursive image methods, and transform-based methods are frequently used for simplified scenarios \cite{Zhou21,Hickson09,Monte00,Monte02,Dias15,Ma04,Rubio21,Goldner92,Dudko04}. On the other hand, numerical schemes like finite difference and finite element methods are favored for handling complex boundary conditions and geometries \cite{Yuan22,Hickson09,Rubio21}. Nevertheless, many studies in the literature focus exclusively on steady-state regimes \cite{Umbricht22,Rubio22,Umbricht22b,Umbricht21,Umbricht20}, limiting their applicability to time-dependent phenomena.

One-dimensional models of diffusion in multilayered media are well established \cite{Norouzi13, Delouei12, Kim20}, and further complexity has been introduced by incorporating spatial or temporal variability in boundary conditions, chemical reactions, and convective terms \cite{Jain09,Yang17,Singh13,Kukla13,Qian17,Kayhani12}. However, in many practical situations, thermal processes must be examined in two or three dimensions, particularly when the geometry or boundary effects play a significant role. Although two-dimensional heat transfer has been previously explored under idealized boundary conditions such as isothermal or adiabatic surfaces orthogonal to the layering direction \cite{ Ma04,Annasabi20,Zukauskas85,Sarkar14,Krishnan23}, these works often exclude the full spectrum of realistic influences.

The existing literature, though rich in scope, still lacks comprehensive models that simultaneously account for multiple effects: transient behavior, convection, internal and external heat generation, and thermal resistance at interfaces. For instance, the work in \cite{Krishnan22} considers two-dimensional heat transfer in layered materials but omits the role of advection, interface resistance, external sources, and general convective boundary conditions—factors that may substantially alter system behavior.

To bridge this gap, the present study develops a detailed mathematical model for unsteady two-dimensional heat transfer in a bilayer structure, governed by a convection-diffusion-reaction-source (CDRS) equation. The formulation incorporates temperature-dependent internal heat generation, advective transport, external heat sources, thermal resistance at the material interface, and general convective conditions on the domain boundaries. Nonlinear effects may play a significant role and, depending on the materials composing the bilayer, can substantially alter its thermal behavior. Therefore, it is important to investigate such effects in specific cases (see, for instance, \cite{Xu24,Xu24b}). In this study, only linear effects at the interface were considered, as incorporating nonlinearities into a general framework would add complexity and could hinder physical interpretation. An analytical solution is constructed using Fourier methods, and a robust finite difference scheme is implemented to simulate specific scenarios. The results aim to enhance theoretical insights and provide useful tools for the design and optimization of thermally complex multilayer systems.

\section{Mathematical Modeling}
\label{Modelo}

The problem of interest concerns transient heat transfer in a two-dimensional bilayer body. Each layer is assumed to be homogeneous and isotropic. Additionally, internal heat generation or loss proportional to the local temperature, as well as advection due to fluid flow, are considered. External heat sources are also included in the model. Phenomena such as thermal runaway and radiative heat transfer are neglected.

\begin{figure}[h!]
\begin{center}
\includegraphics[width=1.00\textwidth]{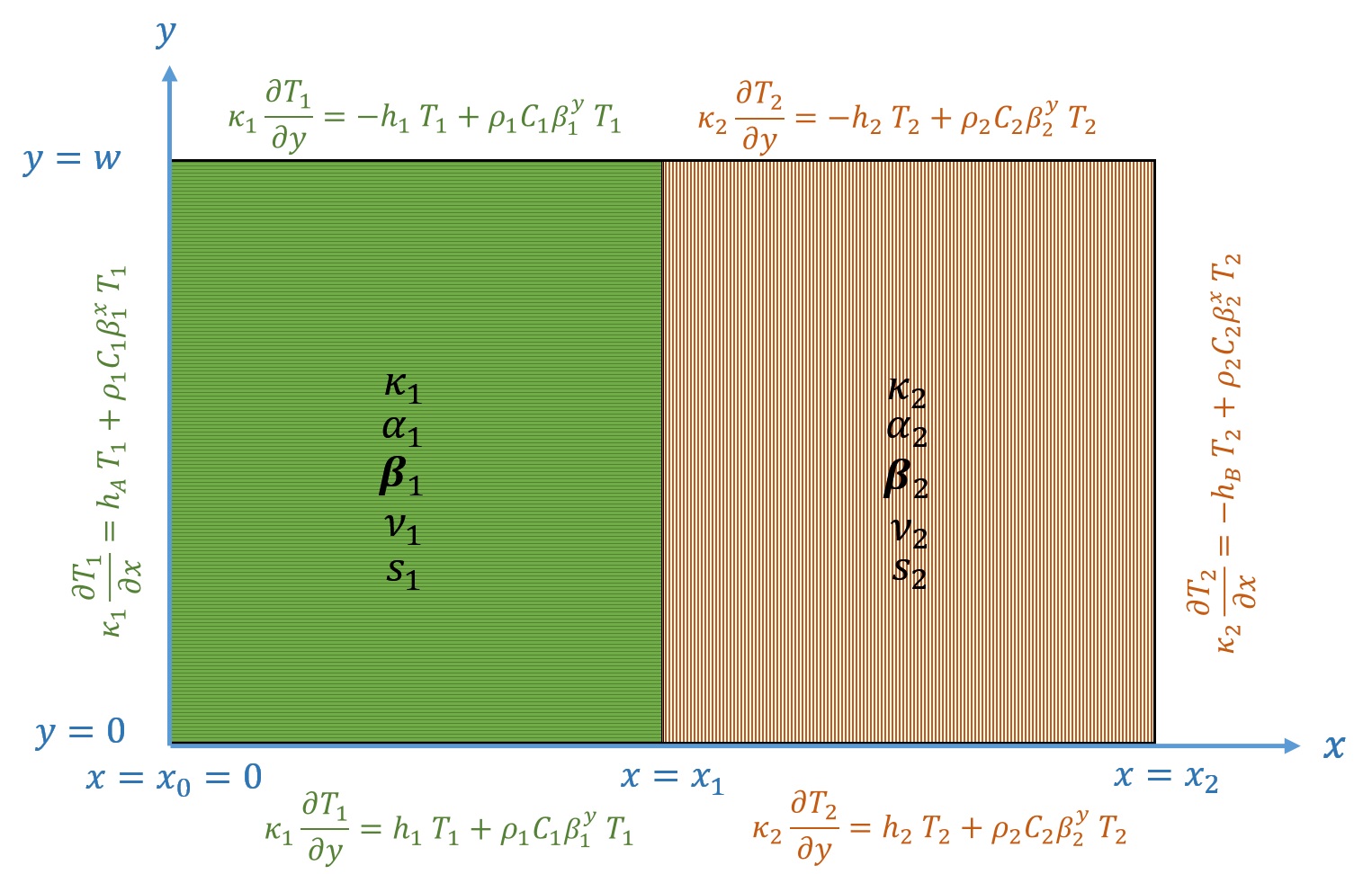}
\caption{General scheme of the problem of interest.}
\label{Esq_Gral}
\end{center}
\end{figure}

The length of the bilayer body is denoted by $x_2$. The interface between the two materials is located at $x_1$, where $x_2 > x_1$. The thickness of the body in the $y$-direction is denoted by $w$. A schematic representation of the problem is shown in Fig. \ref{Esq_Gral}.

The transient energy conservation equation, which accounts for diffusion, advection, internal heat generation or loss, and external heat sources in a two-dimensional bilayer body, is given by:

\begin{equation}
\label{Ec_Parab}
\begin{cases}
\dfrac{\partial{T_1}}{\partial{t}}(x,y,t)= D_1 T_1(x,y,t) + s_1(x,y,t), & \quad (x,y,t) \in \Omega_1,  \\
\\
\dfrac{\partial{T_2}}{\partial{t}}(x,y,t)=D_2 T_2(x,y,t) + s_2(x,y,t), & \quad (x,y,t) \in \Omega_2, 
\end{cases}
\end{equation}
where, for $m=1,2$, $\Omega_m=(x_{m-1},x_m) \times (0,w) \times \R^+$, and $D_m$ is a parabolic differential operator previously used in \cite{Umbricht21b}, defined as
\begin{equation}
\label{Operator_D}
D_m T_m(x,y,t):=  \alpha_{m} \, \Delta \left(T_m\right)(x,y,t) -\bbeta_{m} \cdot \nabla \left(T_m\right)(x,y,t)+ \nu_m \, T_m(x,y,t), 
\end{equation}
where
\begin{equation}
\label{LaplacyGrad}
\begin{cases}
\Delta \left(T_m\right)(x,y,t):= \dfrac{\partial^2{T_m(x,y,t)}}{\partial{x}^2}+ \dfrac{\partial^2{T_m(x,y,t)}}{\partial{y}^2}, \vspace{0.2cm}
\\
\nabla \left(T_m\right)(x,y,t):=\left(\dfrac{\partial{T_m(x,y,t)}}{\partial{x}},\dfrac{\partial{T_m(x,y,t)}}{\partial{y}}\right). 
\end{cases}
\end{equation}

In equations \eqref{Ec_Parab}-\eqref{LaplacyGrad}, often referred to as the CDRS equation, the subscripts denote the first and second material layers, and the variables $x$, $y$, and $t$ represent the spatial and temporal coordinates, respectively. The functions $T_1(x, y, t)$ and $T_2(x, y, t)$, satisfying $T_1(x, y, t) \in C^2(0,x_1) \times C^2(0,w) \times C^1(0,+\infty)$ and $T_2(x, y, t) \in C^2(x_1,x_2) \times C^2(0,w) \times C^1(0,+\infty)$, represent the temperature fields in the first and second layers relative to the ambient temperature at position $(x, y)$ and time $t$.

The first two terms on the right-hand side of equation \eqref{Operator_D} represent heat transfer by diffusion and advection, respectively, while the third term accounts for internal heat generation or loss proportional to the local temperature. The coefficient $\alpha_m$ denotes the thermal diffusivity in each layer, $\bbeta_m=(\beta_m^x,\beta_m^y)$ represents the flow velocity vector, and $\nu_m$ is the coefficient associated with internal heat gain or loss. Finally, the differentiable functions $s_m$ in \eqref{Ec_Parab} model external heat sources. All material properties are assumed to be temperature-independent.

Heat is generated by external sources and also by internal mechanisms within each layer, at a rate proportional to the local temperature. Heat transfer occurs through diffusion and advection due to an imposed fluid flow directed from left to right and from bottom to top. Each layer has its own thermal properties, flow velocity, and heat generation characteristics.

General convective boundary conditions are imposed along the entire boundary of the domain. These conditions represent a balance between convective heat exchange with the environment and internal diffusion and advection. Notably, advection may either supply energy to the body or remove it from the body to the surroundings. The boundary conditions are illustrated in Fig. \ref{Esq_Gral} and are given by (for simplicity, the functional dependence of $T_m$ is omitted below):

\begin{equation}
\label{Cond_Borde}
\begin{cases}
\kappa_1 \, \dfrac{\partial{T_1}}{\partial{x}}=h_A \,T_1 + \rho_1 \, C_1 \, \beta_1^x \,T_1 , & \quad x=0, \,\,\, y \in (0,w), \,\,\,t \in \R^+, 
\vspace{0.2cm}
\\
\kappa_2 \, \dfrac{\partial{T_2}}{\partial{x}}=-h_B \,T_2 + \rho_2 \, C_2  \,\beta_2^x \,T_2 , & \quad x=x_2, \,\,\, y \in (0,w), \,\,\,t \in \R^+, 
\vspace{0.2cm}
\\
\kappa_1 \, \dfrac{\partial{T_1}}{\partial{y}}=h_1 \,T_1 + \rho_1 \, C_1 \, \beta_1^y \,T_1 , & \quad y=0, \,\,\, x \in (0,x_1), \,\,\, t \in \R^+, 
\vspace{0.2cm}
\\
\kappa_1 \, \dfrac{\partial{T_1}}{\partial{y}}=-h_1 \,T_1 + \rho_1 \, C_1 \, \beta_1^y \,T_1 , & \quad y=w, \,\,\, x \in (0,x_1), \,\,\, t \in \R^+, 
\vspace{0.2cm}
\\
\kappa_2 \, \dfrac{\partial{T_2}}{\partial{y}}=h_2 \,T_2 + \rho_2 \, C_2 \, \beta_2^y \,T_2 , & \quad y=0, \,\,\, x \in (x_1,x_2), \,\,\, t \in \R^+, 
\vspace{0.2cm}
\\
\kappa_2 \, \dfrac{\partial{T_2}}{\partial{y}}=-h_2 \,T_2 + \rho_2 \, C_2 \, \beta_2^y \,T_2 , & \quad y=w, \,\,\, x \in (x_1,x_2), \,\,\, t \in \R^+, 
\end{cases} 
\end{equation}
where $\kappa_m$, $h_m$, $\rho_m$, and $C_m$ represent the thermal conductivity, convective heat transfer coefficient, density, and specific heat at constant pressure for each layer, respectively.

The term $h \, T$ in the boundary conditions given in \eqref{Cond_Borde}, corresponds to the classical convective contribution, where the surrounding medium is assumed to be at zero reference temperature. The additional term $\rho C \beta T$ represents an advective heat flux associated with mass transport in the normal direction to the boundary.

A temperature discontinuity at the interface due to thermal contact resistance, along with continuity of the heat flux, is also considered:
\begin{equation}
\label{Cond_Interf}
\begin{cases}
T_2=T_1+ R \, \dfrac{\partial{T_1}}{\partial{x}}, &  x=x_1, \,\, y \in (0,w), \,\, t \in \R^+,  \\
\kappa_2 \, \dfrac{\partial{T_2}}{\partial{x}}- \rho_2 \, C_2 \, \beta_2^x \,T_2=\kappa_1 \, \dfrac{\partial{T_1}}{\partial{x}}- \rho_1 \, C_1 \, \beta_1^x\, T_1, &  x=x_1, \,\, y \in (0,w), \,\, t \in \R^+, \\
\beta_1^y\, \alpha_2=\beta_2^y\, \alpha_1, &  x=x_1, \,\, y \in (0,w), \,\, t \in \R^+, 
\end{cases} 
\end{equation}
the first equation in \eqref{Cond_Interf} models the thermal jump caused by contact resistance, where $R$ is related to the thermal resistance at the material junction which is proportional to the thermal conductivity. This discontinuity is commonly described such that the temperature difference across the interface is proportional to the heat flux \cite{Zhou21b}. The second equation imposes heat flux continuity, consistent with energy conservation. The third interface condition can be interpreted as requiring that the relative balance between advection and diffusion along the $y$-axis is the same in both layers \cite{Cengel07}.

Finally, initial temperature distributions are prescribed for each layer as follows:

\begin{equation}
\label{Cond_Inicial}
\begin{cases}
T_1(x,y,t)=T_{1,0} (x,y),  & \qquad  x \in \left[0,x_1\right], \,\,\,  y \in \left[0,w\right], \,\,\, t=0,  \\
T_2(x,y,t)=T_{2,0} (x,y), & \qquad x \in \left[x_1,x_2\right], \,\,\, y \in \left[0,w\right], \,\,\, t=0. 
\end{cases} 
\end{equation} 

In the next section, we derive an explicit analytical solution to the problem described by equations \eqref{Ec_Parab}-\eqref{Cond_Inicial}.

\section{Analytical Solution} \label{Solucion_analitica}

The transient heat transfer problem to be solved is defined by equations \eqref{Ec_Parab}-\eqref{Cond_Inicial}. For simplicity, the equations are nondimensionalized by introducing the following parameters for $m=1,2$
\begin{equation}
\label{Aux0}
\begin{cases} 
\bar{x}=\dfrac{x}{x_2}, \quad  \bar{y}=\dfrac{y}{x_2}, \quad \bar{w}=\dfrac{w}{x_2}, \quad \bar{R}=\dfrac{R}{x_2}, \quad \tau=\dfrac{\alpha_2}{x_2^2} \,t, \quad \theta_m=\dfrac{T_m}{T_{r}}, \quad \bar{\alpha}=\dfrac{\alpha_1}{\alpha_2}, 
\vspace{0.2cm}\\  {\bPe}_m=\dfrac{x_2}{\alpha_2} \, \bbeta_m, \quad  \bar{\nu}_m=\dfrac{x_2^2}{\alpha_2} \,\nu_m, \quad \bar{s}_m= \dfrac{x_2^2}{T_{r} \, \alpha_2} \, s_m, \quad \bar{\kappa}=\dfrac{\kappa_1}{\kappa_2}, \quad {Bi}_m=\dfrac{x_2}{\kappa_2} \, h_m, 
\vspace{0.2cm}\\  {Bi}_A=\dfrac{x_2}{\kappa_2} \, h_A, \quad {Bi}_B=\dfrac{x_2}{\kappa_2} \, h_B.
\end{cases}
\end{equation}
where $\bPe_m = (Pe_m^{\bar{x}}, Pe_m^{\bar{y}})$ and $Bi_m$ correspond to the dimensionless Péclet and Biot numbers, respectively, and $T_r$ denotes a chosen reference temperature. Applying this nondimensional transformation to equations \eqref{Ec_Parab}–\eqref{Cond_Inicial} yields the following dimensionless formulation of the problem:
\begin{equation}
\label{SisT_adimen}
\begin{cases} 
\dfrac{\partial{\theta_1}}{\partial{\tau}}= \bar{D}_1 \, \theta_1  + \bar{s}_1, \,\,\, & (\bar{x},\bar{y},\tau) \in \bar{\Omega}_1, \vspace{0.2cm}\\
\dfrac{\partial{\theta_2}}{\partial{\tau}}= \bar{D}_2 \, \theta_2  + \bar{s}_2, \,\,\, & (\bar{x},\bar{y},\tau) \in \bar{\Omega}_2,  \vspace{0.2cm}\\
\dfrac{\partial{\theta_1}}{\partial{\bar{x}}}=  {{Bi}_A}^* \,\theta_1 , \,\,\, & \bar{x}=0, \,\,\, \bar{y} \in (0,\bar{w}), \,\,\, \tau \in \R^+, \vspace{0.2cm}\\
\dfrac{\partial{\theta_2}}{\partial{\bar{x}}}=  {{Bi}_B}^* \, \theta_2 , \,\,\, & \bar{x}=1, \,\,\, \bar{y} \in (0,\bar{w}), \,\,\, \tau \in \R^+, \vspace{0.2cm}\\
\dfrac{\partial{\theta_1}}{\partial{\bar{y}}}=  {{Bi}_{1,0}}^* \, \theta_1 ,\,\,\, & \bar{y}=0, \,\,\, \bar{x} \in (0,\bar{x}_1), \,\,\, \tau \in \R^+, \vspace{0.2cm}\\
\dfrac{\partial{\theta_1}}{\partial{\bar{y}}}=  {{Bi}_{1,\bar{w}}}^* \, \theta_1 , \,\,\, & \bar{y}=\bar{w},\,\,\, \bar{x} \in (0,\bar{x}_1), \,\,\, \tau \in \R^+, \vspace{0.2cm}\\
\dfrac{\partial{\theta_2}}{\partial{\bar{y}}}=  {{Bi}_{2,0}}^* \, \theta_2 , \,\,\, & \bar{y}=0, \,\,\, \bar{x} \in (\bar{x}_1,1), \,\,\, \tau \in \R^+, \vspace{0.2cm}\\
\dfrac{\partial{\theta_2}}{\partial{\bar{y}}}=  {{Bi}_{2,\bar{w}}}^* \, \theta_2 , \,\,\, & \bar{y}=\bar{w}, \,\,\, \bar{x} \in (\bar{x}_1,1), \,\,\, \tau \in \R^+, \vspace{0.2cm}\\
\theta_2 = \theta_1 + \bar{R} \, \dfrac{\partial{\theta_1}}{\partial{\bar{x}}} , \,\,\, & \bar{x}=\bar{x}_1,  \,\,\, \bar{y} \in (0,\bar{w}), \,\,\, \tau \in \R^+, \vspace{0.2cm}\\
\dfrac{\partial{\theta_2}}{\partial{\bar{x}}}= \gamma \, \theta_1 + \sigma \, \dfrac{\partial{\theta_1}}{\partial{\bar{x}}}, \,\,\, & \bar{x}=\bar{x}_1,  \,\,\, \bar{y} \in (0,\bar{w}), \,\,\, \tau \in \R^+, \vspace{0.2cm}\\
Pe_1^{\bar{y}}=\bar{\alpha} \, Pe_2^{\bar{y}}, \,\,\, & \bar{x}=\bar{x}_1,  \,\,\, \bar{y} \in (0,\bar{w}), \,\,\, \tau \in \R^+, \vspace{0.2cm}\\
\theta_1=\theta_{1,0}, \,\,\, & \bar{x} \in \left[0,\bar{x}_1\right], \,\,\, \bar{y} \in \left[0,\bar{w}\right], \,\,\, \tau=0,\vspace{0.2cm}\\
\theta_2=\theta_{2,0},\,\,\,  & \bar{x} \in \left[\bar{x}_1,1\right], \,\,\, \bar{y} \in \left[0,\bar{w}\right], \,\,\,\tau=0,
\end{cases}
\end{equation}
where for $m=1,2$; $\bar{\Omega}_m=(\bar{x}_{m-1},\bar{x}_m) \times (0,\bar{w}) \times \R^+$ and $\bar{D}_m$ is the dimensionless parabolic differential operator defined as follows:
\begin{equation}
\label{Oper_adimen}
\begin{cases} 
\bar{D}_1 \theta_1 =\bar{\alpha} \, \Delta \theta_1 - {\bPe}_1 \cdot \nabla \theta_1+ \bar{\nu}_1 \, \theta_1, \vspace{0.2cm} \\
\bar{D}_2 \theta_2 = \Delta \theta_2 - {\bPe}_2 \cdot \nabla \theta_2+ \bar{\nu}_2 \, \theta_2,
\end{cases}
\end{equation}
and 
\begin{equation}
\label{Par_nuevos}
\begin{cases} 
{{Bi}_A}^*=\dfrac{Pe_1^{\bar{x}}}{\bar{\alpha}}+\dfrac{{Bi}_A}{\bar{\kappa}},  \quad
{{Bi}_B}^*= Pe_2^{\bar{x}} - {Bi}_B,\quad 
{{Bi}_{1,0}}^*=\dfrac{Pe_1^{\bar{y}}}{\bar{\alpha}}+\dfrac{{Bi}_1}{\bar{\kappa}}, \vspace{0.2cm}\\ 
{{Bi}_{1,\bar{w}}}^*=\dfrac{Pe_1^{\bar{y}}}{\bar{\alpha}}-\dfrac{{Bi}_1}{\bar{\kappa}}, \quad
{{Bi}_{2,0}}^*= Pe_2^{\bar{y}} + {Bi}_2,\quad 
{{Bi}_{2,\bar{w}}}^*= Pe_2^{\bar{y}} - {Bi}_2,\vspace{0.2cm}\\ 
\gamma= Pe_2^{\bar{x}}- Pe_1^{\bar{x}} \, \dfrac{\bar{\kappa}}{\bar{\alpha}},\quad
\sigma=\bar{\kappa} + \bar{R}\,Pe_2^{\bar{x}}.
\end{cases}
\end{equation}

To simplify equation \eqref{Oper_adimen}, the advective term is removed through a variable substitution that effectively corresponds to shifting the reference frame to one moving with the fluid. This transformation is analogous to those employed in previous studies for tackling similar transport problems (e.g., \cite{Basha93, Bharati17, Das17, Sanskrityayn17}). The change of variables considered here is given by:
\begin{equation}
\label{Def_w}
\begin{cases} 
\theta_1=\exp\left(\bchi_1 \cdot (\bar{x},\bar{y})\right) \, \Theta_1,  \qquad  & (\bar{x},\bar{y},\tau) \in [0,\bar{x}_1] \times [0,\bar{w}] \times \R^+, \vspace{0.2cm}\\
\theta_2=\exp\left(\bchi_2\cdot (\bar{x},\bar{y})\right) \, \Theta_2,  \qquad  & (\bar{x},\bar{y},\tau) \in [\bar{x}_1,1] \times [0,\bar{w}] \times \R^+,
\end{cases}
\end{equation}
where
\begin{equation}
\label{chis}
\bchi_1= \dfrac{1}{2 \,\bar{\alpha}} \, \bPe_1, \qquad \bchi_2= \dfrac{1}{2 } \, \bPe_2,
\end{equation}
are auxiliary parameters included in order to simplify the notation. Applying the transformations defined in \eqref{Def_w}–\eqref{chis} to the system \eqref{SisT_adimen}–\eqref{Par_nuevos} yields the following set of equations:

\begin{equation}
\label{Sist_w1} 
\begin{cases} 
\dfrac{\partial{\Theta_1}}{\partial{\tau}}=\bar{\alpha} \, \Delta \Theta_1  + \psi_1 \, \Theta_1 + \widehat{s}_1, \, & (\bar{x},\bar{y},\tau) \in  \bar{\Omega}_1, \vspace{0.2cm}\\
\dfrac{\partial{\Theta_2}}{\partial{\tau}}= \Delta \Theta_2  + \psi_2 \, \Theta_2 + \widehat{s}_2, \, & (\bar{x},\bar{y},\tau) \in  \bar{\Omega}_2,\vspace{0.2cm}\\
\dfrac{\partial{\Theta_1}}{\partial{\bar{x}}}=  \bar{{Bi}}_A \, \Theta_1 , \,\,\, & \bar{x}=0 ,  \,\,\,\bar{y} \in (0,\bar{w}) ,  \,\,\, \tau \in \R^+, \vspace{0.2cm}\\
\dfrac{\partial{\Theta_2}}{\partial{\bar{x}}}=  \bar{{Bi}}_B \, \Theta_2 ,  \,\,\, & \bar{x}=1 ,  \,\,\, \bar{y} \in (0,\bar{w}) ,  \,\,\, \tau \in \R^+, \vspace{0.2cm}\\
\dfrac{\partial{\Theta_1}}{\partial{\bar{y}}}=  \bar{{Bi}}_{1,0} \, \Theta_1 ,  \,\,\, & \bar{y}=0 ,  \,\,\, \bar{x} \in (0,\bar{x}_1) ,  \,\,\, \tau \in \R^+, \vspace{0.2cm}\\
\dfrac{\partial{\Theta_1}}{\partial{\bar{y}}}=  \bar{{Bi}}_{1,\bar{w}} \, \Theta_1 ,  \,\,\, & \bar{y}=\bar{w} ,  \,\,\, \bar{x} \in (0,\bar{x}_1) ,  \,\,\, \tau \in \R^+, \vspace{0.2cm}\\
\dfrac{\partial{\Theta_2}}{\partial{\bar{y}}}=  \bar{{Bi}}_{2,0} \, \Theta_2 ,  \,\,\, & \bar{y}=0 ,  \,\,\, \bar{x} \in (\bar{x}_1,1) ,  \,\,\, \tau \in \R^+, \vspace{0.2cm}\\
\dfrac{\partial{\Theta_2}}{\partial{\bar{y}}}=  \bar{{Bi}}_{2,\bar{w}} \, \Theta_2 ,  \,\,\, & \bar{y}=\bar{w} ,  \,\,\, \bar{x} \in (\bar{x}_1,1) ,  \,\,\, \tau \in \R^+, \vspace{0.2cm}\\
\Theta_2 = \phi  \, \Theta_1 + \mu  \, \dfrac{\partial{\Theta_1}}{\partial{\bar{x}}} ,  \,\,\, & \bar{x}=\bar{x}_1 ,  \,\,\, y \in (0,\bar{w}) ,  \,\,\,\tau \in \R^+, \vspace{0.2cm}\\
\dfrac{\partial{\Theta_2}}{\partial{\bar{x}}}= \eta  \, \Theta_1 + \varphi  \, \dfrac{\partial{\Theta_1}}{\partial{\bar{x}}},  \,\,\, & \bar{x}=\bar{x}_1 ,  \,\,\, y \in (0,\bar{w}) ,  \,\,\,\tau \in \R^+, \vspace{0.2cm}\\
\Theta_1=\Theta_{1,0}, \,\,\, & \bar{x} \in \left[0,\bar{x}_1\right], \,\,\, \bar{y} \in \left[0,\bar{w}\right],  \,\,\, \tau=0,\vspace{0.2cm}\\
\Theta_2=\Theta_{2,0} , \,\,\,& \bar{x} \in \left[\bar{x}_1,1\right],  \,\,\, \bar{y} \in \left[0,\bar{w}\right],  \,\,\, \tau=0,
\end{cases}
\end{equation}
where
\begin{equation}
\label{Aux1}
\begin{cases} 
\psi_1= \bar{\nu}_1 - \bar{\alpha} \, \left\|\bchi_1\right\|^2, \quad \psi_2= \bar{\nu}_2 - \left\|\bchi_2\right\|^2,  \quad \widehat{s}_1= \bar{s}_1 \, \exp\left(-\bchi_1 \cdot (\bar{x},\bar{y}) \right),\vspace{0.2cm}\\
\widehat{s}_2= \bar{s}_2 \, \exp\left(-\bchi_2 \cdot (\bar{x},\bar{y}) \right), \quad
\bar{{Bi}}_A = {{Bi}_A}^*-\chi_1^{\bar{x}}, \quad \bar{{Bi}}_B = {{Bi}_B}^*-\chi_2^{\bar{x}}, \vspace{0.2cm}\\ \bar{{Bi}}_{1,0} = {{Bi}_{1,0}}^*-\chi_1^{\bar{y}}, \quad \bar{{Bi}}_{1,\bar{w}} = {{Bi}_{1,\bar{w}}}^*-\chi_1^{\bar{y}}, \quad \bar{{Bi}}_{2,0} = {{Bi}_{2,0}}^*-\chi_2^{\bar{y}}, \vspace{0.2cm}\\
\bar{{Bi}}_{2,\bar{w}} = {{Bi}_{2,\bar{w}}}^*-\chi_2^{\bar{y}} \quad 
 \phi = \delta \,\xi  , \quad \mu = \bar{R} \, \xi,\quad \eta =\xi \, \left(\gamma+ \sigma\, \chi_1^{\bar{x}} - \delta\,\chi_2^{\bar{x}}   \right), 
\vspace{0.2cm}\\
\varphi=\xi  \, \left(\sigma - \bar{R} \,\chi_2^{\bar{x}} \right), \quad  \delta=1+ \bar{R} \, \chi_1^{\bar{x}}, \quad  \xi = \exp\left(\frac{1}{2} \, \bar{x}_1(\chi_1^{\bar{x}}-\chi_2^{\bar{x}}) \right), \vspace{0.2cm}\\   \Theta_{1,0} =\theta_{1,0} \, \exp\left(-\bchi_1 \cdot (\bar{x},\bar{y}) \right), \quad \Theta_{2,0} =\theta_{2,0} \, \exp\left(-\bchi_2 \cdot (\bar{x},\bar{y}) \right).

\end{cases}
\end{equation}

We examine the homogeneous counterpart of equations \eqref{Sist_w1}-\eqref{Aux1}, omitting the source terms $\widehat{s}_1$ and $\widehat{s}_2$. This system is addressed using the method of separation of variables. To that end, we assume the existence of functions $f_{1,n,p} \in C^2 (0,\bar{x}_1)$, $f_{2,n,p} \in C^2 (\bar{x}_1,1)$, $u_{1,p} \in C^2 (0,\bar{w})$, $u_{2,p} \in C^2 (0,\bar{w})$ and $g_{n,p} \in C^1 (0,+\infty)$ satisfying the following relation: 
\begin{equation}
\label{SepVar1}
\begin{cases} 
\Theta_1^H(\bar{x},\bar{y},\tau)= \ds \sum_{n=1}^{\infty}\ds \sum_{p=1}^{\infty} {f_{1,n,p} (\bar{x}) \, u_{1,p}(\bar{y}) \, g_{n,p}(\tau)}, \, & (\bar{x},\bar{y},\tau) \in \bar{\Omega}_1,\\ 
\Theta_2^H(\bar{x},\bar{y},\tau)= \ds \sum_{n=1}^{\infty}\ds \sum_{p=1}^{\infty} {f_{2,n,p} (\bar{x}) \, u_{2,p}(\bar{y}) \, g_{n,p}(\tau)},  \, & (\bar{x},\bar{y},\tau) \in \bar{\Omega}_2.
\end{cases}
\end{equation}

Upon substituting expression \eqref{SepVar1} into the homogeneous system derived from equations \eqref{Sist_w1}–\eqref{Aux1}, one obtains that the temporal component satisfies $g_{n,p}(\tau) = K_{n,p} , \exp(-\lambda_{n,p}^2 , \tau)$, where $\lambda_{n,p}$ are the corresponding temporal eigenvalues and $K_{n,p}$ denotes a sequence determined by the initial temperature distribution. Moreover, the spatial functions $f_{m,n,p}$ and $u_{m,p}$, for $m=1,2$, fulfill the following conditions:
\begin{equation}
\label{fs}
\begin{cases} 
\bar{\alpha} \, \left(f''_{1} \, u_{1}+f_{1} \, u''_{1}\right) +\psi_1 \,f_{1}\, u_{1} = - \lambda^2 \, f_{1}\, u_{1}, \quad & (\bar{x},\bar{y}) \in (0,\bar{x}_1) \times (0,\bar{w}), \\ 
f''_{2} \, u_{2}+f_{2} \, u''_{2} +\psi_2 \,f_{2} \, u_{2} = - \lambda^2 \, f_{2}\, u_{2}, \quad & (\bar{x},\bar{y}) \in (\bar{x}_1,1) \times (0,\bar{w}), \\  
f'_{1}=  \bar{{Bi}}_A \, f_{1}, \, & \bar{x}=0 , \\
f'_{2}=  \bar{{Bi}}_B \, f_{2}, \, & \bar{x}=1 , \\
u'_{1}=  \bar{{Bi}}_{1,0} \, u_{1}, \, & \bar{y}=0 , \\
u'_{1}=  \bar{{Bi}}_{1,\bar{w}} \, u_{1}, \, & \bar{y}=\bar{w} , \\
u'_{2}=  \bar{{Bi}}_{2,0} \, u_{2}, \, & \bar{y}=0 , \\
u'_{2}=  \bar{{Bi}}_{2,\bar{w}} \, u_{2}, \, & \bar{y}=\bar{w} , \\
f_{2} = \bar{\phi} \, f_{1} + \bar{\mu} \, f'_{1} , \, & \bar{x}=\bar{x}_1, \,\,\, \bar{y} \in (0,\bar{w}), \\
f'_{2} = \bar{\eta} \, f_{1} + \bar{\varphi} \, f'_{1} , \, & \bar{x}=\bar{x}_1, \,\,\, \bar{y} \in (0,\bar{w}), 
\end{cases}
\end{equation}
for the sake of clarity and brevity, the dependence of parameters and functions on $n$, $p$, $\bar{x}$, and $\bar{y}$ has been omitted from the notation. Additionally,
\begin{equation}
\label{par}
\bar{\phi}=\dfrac{\phi}{q}, \qquad  \bar{\mu}=\dfrac{\mu}{q}, \qquad     \bar{\eta}=\dfrac{\eta}{q}, \qquad     \bar{\varphi}=\dfrac{\varphi}{q},
\end{equation}
where $q \in \R\setminus\{0\} $ is defined such that $u_2(\bar{y})=q \, u_1(\bar{y}) $, $\forall \bar{y} \in (0,\bar{w})$. The existence of such a proportionality constant $q$ follows directly from the interface conditions specified in system \eqref{Sist_w1}.

The coupled system \eqref{fs} is then analyzed under the assumption of a non-trivial solution, leading to the following result:
\begin{equation}
\label{us}
\begin{cases} 
u_{1,p}(\bar{y})= A_{1,p}  \cos (\epsilon_{1,p} \,\bar{y}) + B_{1,p}  \sin (\epsilon_{1,p} \,\bar{y}), \qquad & \bar{y} \in [0,\bar{w}], \\  
u_{2,p}(\bar{y})= A_{2,p}  \cos (\epsilon_{2,p} \,\bar{y}) + B_{2,p}  \sin (\epsilon_{2,p} \,\bar{y}), \qquad & \bar{y} \in [0,\bar{w}],
\end{cases}
\end{equation}
where $A_{1,p}=A_{2,p}=1$, $B_{1,p}=\dfrac{\bar{{Bi}}_{1,0}}{\epsilon_{1,p}}$ and $B_{2,p}=\dfrac{\bar{{Bi}}_{2,0}}{\epsilon_{2,p}}$. The parameters $\epsilon_{1,p}$ and $\epsilon_{2,p}$ represent the spatial eigenvalues in the $y$-direction for the first and second layers, respectively. These values are the infinite solutions of the following eigenvalue equations:
\begin{equation}
\label{AutoVal_y}
\tan (\epsilon_{1,p} \, \bar{w})=\dfrac{\epsilon_{1,p} (\bar{{Bi}}_{1,0}-\bar{{Bi}}_{1,\bar{w}})}{\epsilon_{1,p}^2+\bar{{Bi}}_{1,0} \, \bar{{Bi}}_{1,\bar{w}}}, \qquad
\tan (\epsilon_{2,p} \, \bar{w})=\dfrac{\epsilon_{2,p} (\bar{{Bi}}_{2,0}-\bar{{Bi}}_{2,\bar{w}})}{\epsilon_{2,p}^2+\bar{{Bi}}_{2,0} \, \bar{{Bi}}_{2,\bar{w}}}. 
\end{equation}

Similarly, we have that:
\begin{equation}
\label{fs2}
\begin{cases} 
f_{1,n,p}(\bar{x})= A_{1,n,p}  \cos (\omega_{1,n,p} \,\bar{x}) + B_{1,n,p}  \sin (\omega_{1,n,p} \,\bar{x}), \qquad & \bar{x} \in [0,\bar{x}_1], \\  
f_{2,n,p}(\bar{x})= A_{2,n,p}  \cos (\omega_{2,n,p} \,\bar{x}) + B_{2,n,p}  \sin (\omega_{2,n,p} \,\bar{x}), \qquad & \bar{x} \in [\bar{x}_1,1],
\end{cases}
\end{equation}
where $A_{1,n,p}=1$, $B_{1,n,p}=\frac{\bar{{Bi}}_A}{\omega_{1,n,p}}$ and 
\begin{equation}
\label{An}
\begin{split}
A_{2,n,p}=& \dfrac{\sin(\omega_{1,n,p} \, \bar{x}_1)}{\cos(\omega_{2,n,p} \, \bar{x}_1)} \left(\bar{\phi} \, \dfrac{\bar{{Bi}}_A}{\omega_{1,n,p}}-\bar{\mu} \, \omega_{1,n,p} \right)
  + \dfrac{\cos(\omega_{1,n,p} \, \bar{x}_1)}{\cos(\omega_{2,n,p} \, \bar{x}_1)} \left(\bar{\phi} +\bar{\mu} \, \bar{{Bi}}_A \right) - \tan (\omega_{2,n,p} \, \bar{x}_1) \, B_{2,n,p}   
\end{split}
\end{equation}
\begin{equation}
\label{Bn}
\begin{split}
 B_{2,n,p} = & \sin(\omega_{2,n,p} \, \bar{x}_1) \left[\sin(\omega_{1,n,p} \, \bar{x}_1) \left(\bar{\phi} \, \dfrac{\bar{{Bi}}_A}{\omega_{1,n,p}} - \bar{\mu} \, \omega_{1,n,p} \,\right) + \cos(\omega_{1,n,p} \, \bar{x}_1) \left(\bar{\phi} + \bar{\mu} \, \bar{{Bi}}_A \, \right) \right] \\
 +& \dfrac{\cos(\omega_{2,n,p} \, \bar{x}_1)}{\omega_{2,n,p}}  \left[\sin(\omega_{1,n,p} \, \bar{x}_1)\left(\bar{\eta} \,\dfrac{ \bar{{Bi}}_A}{\omega_{1,n,p}} - \bar{\varphi} \, \omega_{1,n,p}\right) + \cos(\omega_{1,n,p} \, \bar{x}_1) \left(\bar{\eta}+\bar{\varphi }\, \bar{{Bi}}_A\right)\right].
\end{split}
\end{equation}

The parameters $\omega_{1,n,p}$ and $\omega_{2,n,p}$ represent the spatial eigenvalues in the $x$-direction for the first and second layers, respectively, and are defined from: 
\begin{equation}
\label{Omegas}
\begin{cases} 
\omega_{1,n,p}=\omega_{1,n,p}(\lambda_{n,p},\epsilon_{1,p})=\sqrt{\dfrac{\lambda_{n,p}^2+\psi_1}{\bar{\alpha}}-\epsilon_{1,p}^2},\vspace{0.1cm} \\ 
\omega_{2,n,p}=\omega_{2,n,p}(\lambda_{n,p},\epsilon_{2,p})=\sqrt{\lambda_{n,p}^2+\psi_2-\epsilon_{2,p}^2},
\end{cases}
\end{equation}
these values are the infinite solutions of the following eigenvalue equations:
\begin{equation}
\label{Tang}
\tan (\omega_{2,n,p})=\dfrac{\omega_{2,n,p} \, B_{2,n,p}- \bar{{Bi}}_B \, A_{2,n,p}}{\bar{{Bi}}_B \, B_{2,n,p}+\omega_{2,n,p}\,  A_{2,n,p}}.
\end{equation}

\begin{nt}
An important point to note is that the spatial eigenvalues in the \( x \) direction (\( \omega_{1,n,p} \), \( \omega_{2,n,p} \)) depend on the temporal eigenvalues (\( \lambda_{n,p} \)) and the spatial eigenvalues in the \( y \) direction (\( \epsilon_{1,p} \), \( \epsilon_{2,p} \)). The existence of infinitely many eigenvalues for these types of problems has already been discussed in \cite{Krishnan22,Jain21, Umbricht25,Umbricht25b}. In the is work, only real eigenvalues will be considered since we assume no superheating or thermal runaway \cite{Esho18} in the thermal process under study.
\end{nt}

To solve the non-homogeneous system defined by \eqref{Sist_w1}-\eqref{Aux1}, the Fourier method is employed on the solution of the homogeneous problem outlined in \eqref{SepVar1}. In particular, it is assumed that there exist two countably infinite sets of time functions,  
$\bar{A}_{n,p}(\tau)$ and $\bar{B}_{n,p}(\tau)$, such that:
\begin{equation}
\label{SolNoHomegeneo2}
\begin{cases} 
\Theta_1(\bar{x},\bar{y},\tau)= \ds \sum_{n=1}^{\infty}\ds \sum_{p=1}^{\infty} \bar{A}_{n,p}(\tau) \, {f_{1,n,p} (\bar{x}) \, u_{1,p}(\bar{y})}, \,\,\, & (\bar{x},\bar{y},\tau) \in [0,\bar{x}_1] \times [0,\bar{w}]\times \R^+, \\  
\Theta_2(\bar{x},\bar{y},\tau)= \ds \sum_{n=1}^{\infty}\ds \sum_{p=1}^{\infty}  \bar{B}_{n,p}(\tau) \, {f_{2,n,p} (\bar{x}) \, u_{2,p}(\bar{y})}, \,\,\, & (\bar{x},\bar{y},\tau) \in [\bar{x}_1,1] \times [0,\bar{w}]\times \R^+,
\end{cases} 
\end{equation}
where the functions \( u_{m,p} \) and \( f_{m,n,p} \), with \( m = 1,2 \), are defined in \eqref{us} and \eqref{fs2}, respectively. For simplicity, the source terms \( \widehat{s}_1(y,\tau) \) and \( \widehat{s}_2(y,\tau) \) from \eqref{Sist_w1} are expanded into series of eigenfunctions.
\begin{equation}
\label{Fuenteenserie}
\begin{cases} 
\widehat{s}_1(\bar{x},\bar{y},\tau)= \ds \sum_{n=1}^{\infty}\ds \sum_{p=1}^{\infty} { S_{1,n,p}(\tau) \, f_{1,n,p} (\bar{x}) \, u_{1,p} (\bar{y}) }, \,\,\, & (\bar{x},\bar{y},\tau) \in [0,\bar{x}_1] \times [0,\bar{w}]\times \R^+,  \\ 
\widehat{s}_2(\bar{x},\bar{y},\tau)= \ds \sum_{n=1}^{\infty}\ds \sum_{p=1}^{\infty} { S_{2,n,p}(\tau) \, f_{2,n,p} (\bar{x}) \, u_{2,p} (\bar{y})  }, \,\,\, & (\bar{x},\bar{y},\tau) \in [\bar{x}_1,1] \times [0,\bar{w}]\times \R^+,
\end{cases} 
\end{equation}
where $S_{1,n,p}(\tau)$ and $S_{2,n,p}(\tau)$ are defined as follows
\begin{equation}
\label{Gs1}
S_{1,n,p}(\tau)=\dfrac{\ds \int_0^{\bar{x}_1} \ds \int_0^{\bar{w}}\widehat{s}_1(\bar{x},\bar{y},\tau) \,  f_{1,n,p} (\bar{x}) \, u_{1,p}(\bar{y}) \, d\bar{y}\,d\bar{x}}{\ds \int_0^{\bar{x}_1} \ds \int_0^{\bar{w}} \left[f_{1,n,p} (\bar{x}) \, u_{1,p}(\bar{y})\right]^2 \, d\bar{y}\,d\bar{x}}, 
\end{equation}
\begin{equation}
\label{Gs2}
S_{2,n,p}(\tau)=\dfrac{\ds \int_{\bar{x}_1}^1 \ds \int_0^{\bar{w}}\widehat{s}_2(\bar{x},\bar{y},\tau) \,  f_{2,n,p} (\bar{x}) \, u_{2,p}(\bar{y}) \, d\bar{x}d\bar{y}}{\ds \int_{\bar{x}_1}^1 \ds \int_0^{\bar{w}} \left[f_{2,n,p} (\bar{x}) \, u_{2,p}(\bar{y})\right]^2 \, d\bar{x}\,d\bar{y}}.
\end{equation}

By inserting expressions \eqref{SolNoHomegeneo2}–\eqref{Gs2} into equation \eqref{Sist_w1}, one arrives at the following countable system of homogeneous ordinary differential equations:
\begin{equation}
\label{EDO}
\begin{cases} 
\ds \sum_{n=1}^{\infty} \sum_{p=1}^{\infty}\left\{\bar{A}_{n,p}'(\tau) +\left[\bar{\alpha} \, (\omega^2_{1,n,p}+\epsilon^2_{1,p}) - \psi_1\right] \, \bar{A}_{n,p}(\tau) -S_{1,n,p}(\tau)\right\} =0, \\
\ds \sum_{n=1}^{\infty} \sum_{p=1}^{\infty}\left\{\bar{B}_{n,p}'(\tau) +\left[\omega^2_{2,n,p}+\epsilon^2_{2,p} - \psi_2\right] \, \bar{B}_{n,p}(\tau) -S_{2,n,p}(\tau)\right\} =0,
\end{cases}
\end{equation}

Given that eigenfunction expansions in linear systems exhibit properties analogous to those of Fourier series, the vanishing of the series in \eqref{EDO} implies that each individual term must vanish. This requirement leads to a solvable system through direct integration, yielding:
\begin{equation}
\label{solEDO}
\begin{cases} 
\bar{A}_{n,p}(\tau)=\exp\left((\psi_1-\bar{\alpha} (\omega^2_{1,n,p}+\epsilon^2_{1,p}))\, \tau\right) \left[K_{n,p} + \ds \int_0^\tau   S_{1,n,p} (s) \, \exp\left((\bar{\alpha} (\omega^2_{1,n,p}+\epsilon^2_{1,p})-\psi_1)\, s\right)  \,ds\right] , \\
\bar{B}_{n,p}(\tau)=\exp\left((\psi_2-\omega^2_{2,n,p}-\epsilon^2_{2,p})\, \tau\right) \left[K_{n,p} + \ds \int_0^\tau   S_{2,n,p} (s) \, \exp\left((\omega^2_{2,n,p}+\epsilon^2_{2,p}-\psi_2)\, s\right)  \,ds\right].
\end{cases}
\end{equation}

The only remaining unknown is the sequence \( K_{n,p} \), which is determined by enforcing the initial conditions of system \eqref{Sist_w1} and employing the orthogonality property of the eigenfunctions. The procedure for this will be outlined in Section \ref{Ortogonalidad}. As a result, we obtain:
\begin{equation}
\label{CI}
K_{n,p}=\dfrac{\dfrac{\varphi \,\phi - \eta \,\mu}{\bar{\alpha}} \, \ds \int_0^{\bar{x}_1} \int_0^{\bar{w}} \Theta_{1,0} f_{1,n,p}(\bar{x}) u_{1,p}(\bar{y}) \, d\bar{y} \, d\bar{x} +\int_{\bar{x}_1}^1 \int_0^{\bar{w}} \Theta_{2,0}\, f_{2,n,p}(\bar{x}) u_{2,p}(\bar{y})  \, d\bar{y} \, d\bar{x}}{\dfrac{\varphi \,\phi - \eta \,\mu}{\bar{\alpha}} \, \ds\int_0^{\bar{x}_1} \int_0^{\bar{w}}[f_{1,n,p}(\bar{x}) u_{1,p}(\bar{y})]^2 \, d\bar{y} \, d\bar{x} +\int_{\bar{x}_1}^1 \int_0^{\bar{w}} \, [f_{2,n,p}(\bar{x}) u_{2,p}(\bar{y})]^2  \, d\bar{y} \, d\bar{x}}
\end{equation}

\section{Consistency of the solution} \label{Consistencia}

To assess the validity of the analytical solution derived in this work, we compare it with the reference solution presented in \cite{Krishnan22}. That study addresses a similar physical configuration but under a set of simplifying assumptions that are particularly useful for validation purposes. Specifically, the authors consider a case without external heat sources (\( s_1 = s_2 = 0 \)), neglect thermal contact resistance at the interface (\( R = 0 \)), omit advective heat transport (\( \bbeta_{1} = \bbeta_{2} = (0,0) \)), and impose adiabatic boundary conditions (\( h_1 = h_2 = 0 \)). The goal here is to demonstrate that, under these simplifications, both solutions coincide.

These conditions are applied to the analytical solution derived in Section~\ref{Solucion_analitica}. Since \( \bbeta_1 = \bbeta_2 = (0,0) \), it follows that \( \bPe_1 = \bPe_2 = (0,0) \). Likewise, setting \( R = 0 \) implies \( \bar{R} = 0 \), and the absence of internal heat sources (\( s_1 = s_2 = 0 \)) yields \( \bar{s}_1 = \bar{s}_2 = 0 \). Finally, the adiabatic boundary condition (\( h_1 = h_2 = 0 \)) leads to \( Bi_1 = Bi_2 = 0 \).

Under these assumptions, we obtain \( \bchi_1 = \bchi_2 = (0,0) \), from which it follows that \( \theta_1 = \Theta_1 \) and \( \theta_2 = \Theta_2 \). Consequently, the following simplifications hold:
\begin{equation}
\label{Par_nuevos_simp}
\begin{cases} 
{{Bi}_A}^*=\dfrac{{Bi}_A}{\bar{\kappa}}, \quad
{{Bi}_B}^*= - {Bi}_B,\quad 
{{Bi}_{1,0}}^*={{Bi}_{1,\bar{w}}}^*={{Bi}_{2,0}}^*={{Bi}_{2,\bar{w}}}^*=0, \\
\sigma=\varphi=\bar{\kappa},
\Psi_1=\bar{\nu}_1, \quad \Psi_2=\bar{\nu}_2, \quad \widehat{s}_1 = \widehat{s}_2 = 0,
\quad \xi=\delta=\phi=1,\\ \gamma=\eta=\mu=0, 
\quad \bar{Bi}_{1,0}=\bar{Bi}_{1,\bar{w}}=\bar{Bi}_{2,0}=\bar{Bi}_{2,\bar{w}}=0, \quad  \bar{Bi}_A={{Bi}_A}^*, \\
 \bar{Bi}_B={{Bi}_B}^*, \quad  \bar{\mu}=\bar{\eta}=0, \quad \bar{\phi}=\dfrac{1}{q}, \quad \bar{\varphi}=\dfrac{\bar{\kappa}}{q} .
\end{cases}
\end{equation}

On the other hand, from \eqref{Par_nuevos_simp} it follows that:
\begin{equation}
\label{Par_nuevos_simp2}
\epsilon_{1,p}=\epsilon_{2,p}=\epsilon_{p}=\dfrac{n \,\pi }{\bar{w}}, \,\, n \in \N \,\,\, \Rightarrow \,\,\,
u_{1,p}=u_{2,p}=u_{p}=\cos(\epsilon_{p} \, \bar{y}),
\end{equation}
furthermore, using \eqref{Par_nuevos_simp2} we obtain:
\begin{equation}
\label{Omegas_simp}
\omega_{1,n,p}=\sqrt{\dfrac{\lambda_{n,p}^2+\psi_1}{\bar{\alpha}}-\epsilon_{p}^2}, \qquad
\omega_{2,n,p}=\sqrt{\lambda_{n,p}^2+\psi_2-\epsilon_{p}^2}
\end{equation}

and
\begin{equation}
\label{An_simp}
\begin{split}
A_{2,n,p}=& \dfrac{\sin(\omega_{1,n,p} \, \bar{x}_1)}{\cos(\omega_{2,n,p} \, \bar{x}_1)} \left(\bar{\phi} \, \dfrac{\bar{{Bi}}_A}{\omega_{1,n,p}} \, \omega_{1,n,p} \right)
  + \dfrac{\cos(\omega_{1,n,p} \, \bar{x}_1)}{\cos(\omega_{2,n,p} \, \bar{x}_1)} \bar{\phi}
	- \tan (\omega_{2,n,p} \, \bar{x}_1) \, B_{2,n,p}   
\end{split}
\end{equation}
\begin{equation}
\label{Bn__simp}
\begin{split}
 B_{2,n,p} = & \sin(\omega_{2,n,p} \, \bar{x}_1) \left[\sin(\omega_{1,n,p} \, \bar{x}_1) \left(\bar{\phi} \, \dfrac{\bar{{Bi}}_A}{\omega_{1,n,p}}\right) + \cos(\omega_{1,n,p} \, \bar{x}_1) \, \right] \\
 +& \dfrac{\cos(\omega_{2,n,p} \, \bar{x}_1)}{\omega_{2,n,p}}  \left[\sin(\omega_{1,n,p} \, \bar{x}_1)\left( - \bar{\varphi} \, \omega_{1,n,p}\right) + \cos(\omega_{1,n,p} \, \bar{x}_1) \left(\bar{\varphi }\, \bar{{Bi}}_A\right)\right].
\end{split}
\end{equation}

Finally, the equations \eqref{Par_nuevos_simp2}-\eqref{Bn__simp} give rise to:
\begin{equation}
\label{fs2_simp}
\begin{cases} 
f_{1,n,p}(\bar{x})= \cos (\omega_{1,n,p} \,\bar{x}) + \frac{\bar{Bi}_A}{\omega_{1,n,p}} \sin (\omega_{1,n,p} \,\bar{x}), \qquad & \bar{x} \in [0,\bar{x}_1], \\  
f_{2,n,p}(\bar{x})= A_{2,n,p}  \cos (\omega_{2,n,p} \,\bar{x}) + B_{2,n,p}  \sin (\omega_{2,n,p} \,\bar{x}), \qquad & \bar{x} \in [\bar{x}_1,1],
\end{cases}
\end{equation}
By employing the expression \( g_{n,p} (\tau) = K_{n,p} \, \exp(-\lambda_{n,p}^2 \, \tau) \), together with the simplified spatial functions provided in \eqref{fs2_simp}, the analytical solution corresponding to this particular configuration is derived as:
\begin{equation}
\label{SolNoHomegeneo2b}
\begin{cases} 
\theta_1(\bar{x},\bar{y},\tau)= \ds \sum_{n=1}^{\infty}\ds \sum_{p=1}^{\infty} {f_{1,n,p} (\bar{x}) \, u_{1,p}(\bar{y})} \, g_{n,p} (\tau), \,\,\, & (\bar{x},\bar{y},\tau) \in [0,\bar{x}_1] \times [0,\bar{w}]\times \R^+, \\  
\theta_2(\bar{x},\bar{y},\tau)= \ds \sum_{n=1}^{\infty}\ds \sum_{p=1}^{\infty}  {f_{2,n,p} (\bar{x}) \, u_{2,p}(\bar{y})}\, g_{n,p} (\tau), \,\,\, & (\bar{x},\bar{y},\tau) \in [\bar{x}_1,1] \times [0,\bar{w}]\times \R^+,
\end{cases} 
\end{equation}
where
\begin{equation}
\label{CI_simp}
K_{n,p}=\dfrac{\dfrac{\bar{\kappa}}{\bar{\alpha}} \, \ds \int_0^{\bar{x}_1} \int_0^{\bar{w}} \theta_{1,0} f_{1,n,p}(\bar{x}) u_{p}(\bar{y}) \, d\bar{y} \, d\bar{x} +\int_{\bar{x}_1}^1 \int_0^{\bar{w}} \theta_{2,0}\, f_{2,n,p}(\bar{x}) u_{2,p}(\bar{y})  \, d\bar{y} \, d\bar{x}}{\dfrac{\bar{\kappa}}{\bar{\alpha}} \, \ds\int_0^{\bar{x}_1} \int_0^{\bar{w}}[f_{1,n,p}(\bar{x}) u_{p}(\bar{y})]^2 \, d\bar{y} \, d\bar{x} +\int_{\bar{x}_1}^1 \int_0^{\bar{w}} \, [f_{2,n,p}(\bar{x}) u_{p}(\bar{y})]^2  \, d\bar{y} \, d\bar{x}}.
\end{equation}

In summary, when analyzing the solution derived in this work under the assumptions of transient heat transfer with no internal heat sources, no advection, zero thermal contact resistance at the interface, and adiabatic boundary conditions, it is observed that the resulting formulation fulfills the same conditions established by the authors in \cite{Krishnan22}.

\section{Numerical Modelling}

The analytical resolution of this class of problems entails significant computational effort, rendering it impractical for obtaining temperature distributions in particular scenarios. Consequently, numerical approaches are generally preferred, as they enable efficient computation of temperature profiles and extraction of relevant thermal information.

Among the various numerical techniques, the finite difference method (FDM) stands out as a practical and robust strategy for transient heat transfer simulations. When applied to composite or multilayered domains, special care must be taken at material interfaces, especially in cases where temperature continuity is not guaranteed. Some researchers have addressed this challenge by introducing auxiliary or fictitious layers at the interface; see, for instance, \cite{Yuan22}.

In the present study, we introduce an explicit finite difference scheme of second-order accuracy, utilizing a forward time integration and a central differencing approach in space. Boundary treatments are handled with one-sided differences: forward differences are employed at the left boundary, while backward differences are used at the right. At the internal interface, the method alternates between forward and backward spatial differences depending on whether the adjacent material lies to the left or right, respectively.

In order to implement the numerical method, two uniform 3D partitions are defined on the spatial variable \( x \), the spatial variable \( y \), and the temporal variable \( t \), as a discrete set \( \mathcal{P} \) satisfying:
\begin{equation}
\label{particion1}
\begin{cases}
\mathcal{P}_1=\{ (x_i,y_j,t_k)/ \, i=1,2,\ldots,n_{x_1} ;\, j=1,2,\ldots,n_w;\, k=1,2,\ldots,n_{t_\infty};\,
\\ x_i \in \mathcal{P}_x^1,\, y_j \in \mathcal{P}_y, t_k \in \mathcal{P}_t \}, \\
\\
\mathcal{P}_2=\{ (x_i,y_j,t_k)/ \, i=n_{x_1},n_{x_1}+1,...,n_{x_2} ;\, j=1,2,\ldots,n_w;\, k=1,2,\ldots,n_{t_\infty};\,
\\ x_i \in \mathcal{P}_x^2,\, y_j \in \mathcal{P}_y, t_k \in \mathcal{P}_t \},
\end{cases}
\end{equation}
where
\begin{equation}
\label{particion2}
\begin{cases}
\mathcal{P}_x^1=\{ x_1< \cdots < x_i< \cdots <x_{n_{x_1}}, \,\,\, x_i=(i-1) \Delta x, \,\,\, i=1,2,...,n_{x_1}\} \\
\mathcal{P}_x^2=\{ x_{n_{x_1}}<  \cdots < x_i< \cdots <x_{n_{x_2}}, \,\,\, x_i=(i-1) \Delta x, \,\,\, i=n_{x_1}, n_{x_1}+1,...,n_{x_2}\}
\end{cases}
\end{equation}
\begin{equation}
\label{particion3}
\mathcal{P}_y=\{ y_1< \cdots < y_j< \cdots <y_{n_w}, \,\,\, y_j=(j-1) \Delta y, \,\,\, j=1,2,...,n_{w}\} \\
\end{equation}
and
\begin{equation}
\label{particion4}
\mathcal{P}_t=\{ t_1< \cdots < t_k< \cdots <t_{n_{t_\infty}}, \,\,\, t_k=(k-1) \Delta t, \,\,\, k=1, 2,...,n_{t_\infty}\}.
\end{equation}

Specifically, \( \mathcal{P}_x^i \) with \( i = 1, 2 \) represents the partition of the spatial variable associated with \( x \); \( \mathcal{P}_y \) is the partition of the spatial variable associated with \( y \), and \( \mathcal{P}_t \) is the partition of the temporal variable associated with \( t \). The values of \( \Delta x \), \( \Delta y \), and \( \Delta t \) correspond to the spatial and temporal discretization steps, respectively. These values are determined numerically and defined on an equidistant (uniform) grid as \( \Delta x = x_i - x_{i-1} \), \( \Delta y = y_j - y_{j-1} \), and \( \Delta t = t_k - t_{k-1} \).

The following temperature function is considered:
\begin{equation}
\label{upartida}
T(x,y,t)=
\begin{cases}
T_1(x,y,t), & \quad (x,y,t) \in [0,x_1] \times [0,w] \times [0,t_\infty],  \\
T_2(x,y,t), & \quad (x,y,t) \in [x_1,x_2] \times [0,w] \times [0,t_\infty].
\end{cases}
\end{equation}

To compute the numerical solution of the heat transfer problem under consideration, the equations \eqref{Ec_Parab}–\eqref{Cond_Inicial} are discretized according to the described scheme. As a result, an algebraic system corresponding to the differential equations \eqref{Ec_Parab}–\eqref{LaplacyGrad} is obtained.
\begin{equation}
\label{Discret_Equat}
\begin{cases} 
T^1_{i,j,k+1}=\zeta_{11}\, T^1_{i+1,j,k}+\zeta_{12}\, T^1_{i,j+1,k} + \zeta_{13}\, T^1_{i,j,k}+ \zeta_{14}\, T^1_{i-1,j,k} +
\zeta_{15}\, T^1_{i,j-1,k} 
+ P^1_{i,j,k}, \\
i=2,...,n_{x_1}-1, \quad j=2,...,n_w-1, \quad k=2,...,n_{t_\infty}, \vspace{0.1cm}\\

T^2_{i,j,k+1}=\zeta_{21}\, T^2_{i+1,j,k}+\zeta_{22}\, T^2_{i,j+1,k} + \zeta_{23}\, T^2_{i,j,k}+ \zeta_{24}\, T^2_{i-1,j,k} +
\zeta_{25}\, T^2_{i,j-1,k} 
+ P^2_{i,j,k}, \\
i=n_{x_1}+1,...,n_{x_2}-1,\quad j=2,...,n_w-1, \quad k=2,...,n_{t_\infty}, 
\end{cases}
\end{equation}

the boundary conditions specified in system \eqref{Cond_Borde} are likewise discretized, resulting in the following set of algebraic expressions:
\begin{equation}
\label{Discret_Cond_Borde}
\begin{cases} 
T^1_{i,j,k}= \iota_A \, T^1_{i+1,j,k},  \quad i=1, \quad j=2,...,n_w, \quad k=2,...,n_{t_\infty}, \vspace{0.1cm}\\

T^2_{i,j,k}=\iota_B \,T^2_{i-1,j,k},  \quad i=n_{x_2}, \quad j=2,...,n_w, \quad k=2,...,n_{t_\infty}, \vspace{0.1cm}\\

T^1_{i,j,k}= \iota_{1,0} \, T^1_{i,j+1,k},  \quad i=2,...,n_{x_1}-1, \quad j=1, \quad k=2,...,n_{t_\infty}, \vspace{0.1cm}\\

T^1_{i,j,k}= \iota_{1,w} \, T^1_{i,j-1,k},  \quad i=2,...,n_{x_1}-1, \quad j=n_w, \quad k=2,...,n_{t_\infty}, \vspace{0.1cm}\\

T^2_{i,j,k}= \iota_{2,0} \, T^2_{i,j+1,k},  \quad i=n_{x_1}+1,..., n_{x_2}-1 \quad j=1, \quad k=2,...,n_{t_\infty}, \vspace{0.1cm}\\

T^2_{i,j,k}= \iota_{2,w} \, T^2_{i,j-1,k},  \quad i=n_{x_1}+1,..., n_{x_2}-1 \quad j=n_w, \quad k=2,...,n_{t_\infty}. 
\end{cases}
\end{equation}

On the other hand, discretizing the interface conditions presented in \eqref{Cond_Interf} yields the following system of algebraic equations:

\label{Discret_Cond_Interf}
\begin{equation}
\begin{cases} 
T^1_{i,j,k}=\upsilon_{11} T^1_{i-1,j,k}+ \upsilon_{12} T^2_{i+1,j,k}, \quad &  i=n_{x_1}, \quad j=2,...,n_w, \quad k=2,...,n_{t_\infty},\vspace{0.1cm}\\
T^2_{i,j,k}=\upsilon_{13} T^1_{i-1,j,k}+ \upsilon_{14} T^2_{i+1,j,k}, \quad &  i=n_{x_1}, \quad j=2,...,n_w,\quad  k=2,...,n_{t_\infty}.
\end{cases}
\end{equation}

Finally, applying the same discretization scheme to the initial conditions defined in \eqref{Cond_Inicial} results in the following algebraic expressions:
\begin{equation}
\label{Discret_Cond_Inicial}
\begin{cases} 
T^1_{i,j,k}=T^1_{i,j},  \quad i=1,...,n_{x_1}, \,\, j=1,...,n_w, \,\, k=1, \vspace{0.1cm}\\
T^2_{i,j,k}=T^2_{i,j},  \quad i=n_{x_1},...,n_{x_2}, \,\, j=1,...,n_w, \,\, k=1.
\end{cases}
\end{equation}

The equations \eqref{Discret_Equat}–\eqref{Discret_Cond_Inicial} constitute the discrete formulation of the problem under consideration, where for \( m = 1, 2 \):
\begin{equation}
\label{OtroAux}
\begin{cases} 
\zeta_{m1}=\dfrac{\alpha_m \,\Delta t }{(\Delta x)^2}-\dfrac{\beta_m^x \,\Delta t }{2 \, \Delta x}, \quad

\zeta_{m2}=\dfrac{\alpha_m \,\Delta t }{(\Delta y)^2} -\dfrac{\beta_m^y \,\Delta t }{2 \, \Delta y},   \quad \vspace{0.2cm} \\

\zeta_{m3}=1+ \nu_m \,\Delta t - 2 \dfrac{\alpha_m \,\Delta t }{(\Delta x)^2}- 2 \dfrac{\alpha_m \,\Delta t }{(\Delta y)^2} \quad  \vspace{0.2cm} \\

\zeta_{m4}=\dfrac{\alpha_m \,\Delta t }{(\Delta x)^2}+\dfrac{\beta_m^x \,\Delta t }{2 \, \Delta x}, \quad

\zeta_{m5}= \dfrac{\alpha_m \,\Delta t }{(\Delta y)^2} +\dfrac{\beta_m^y \,\Delta t }{2 \, \Delta y},   \quad

P^m_{i,j,k}=s^m_{i,j,k} \, \Delta t. 
\end{cases}
\end{equation}

Additionally, 
\begin{equation}
\label{iotaA_B}
\iota_A =\dfrac{1}{1+ \Pi_A \, \Delta x},  \qquad  \iota_B =\dfrac{1}{1+ \Pi_B \, \Delta x} 
\end{equation}
and for $m=1,2$
\begin{equation}
\label{iotas}
\iota_{m,0} =\dfrac{1}{1+ \Pi_{m,0} \, \Delta y},  \qquad  \iota_{m,w} =\dfrac{1}{1+ \Pi_{m,w} \, \Delta y},  
\end{equation}
where
\begin{equation}
\label{OtroAux2}
\Pi_A=\dfrac{h_A}{\kappa_1}+\dfrac{\beta_1^x}{\alpha_1}, \quad \Pi_B=\dfrac{h_B}{\kappa_2}-\dfrac{\beta_2^x}{\alpha_2},
\quad \Pi_{m,0}=\dfrac{h_m}{\kappa_m}+\dfrac{\beta_m^y}{\alpha_m}, \quad 
\Pi_{m,w}=\dfrac{h_m}{\kappa_m}-\dfrac{\beta_m^y}{\alpha_m}. 
\end{equation}

Finally
\begin{equation}
\label{OtroAux3}
\upsilon_{1,1}=\dfrac{\kappa_1+\kappa_2 \, z_2 \, \Upsilon}{\Lambda}, \quad \upsilon_{1,2}=\dfrac{\kappa_2 }{\Lambda},\quad 
\upsilon_{1,3}=\dfrac{\kappa_1(1+\Upsilon-z_1 \, \Upsilon)}{\Lambda}, \quad  \upsilon_{1,4}=\dfrac{\kappa_2 (1+ \Upsilon) }{\Lambda},
\end{equation}
with
\begin{equation}
\label{OtroAux4}
\Upsilon=\dfrac{R}{\Delta x}, \quad \Lambda=\kappa_1\, z_1 + \kappa_2\, z_2 (1+\Upsilon), \quad 
z_1=1-\dfrac{\beta_1^x}{\alpha_1}\, \Delta x, \quad  z_2=1+\dfrac{\beta_2^x}{\alpha_2}\, \Delta x.
\end{equation}

The collection of equations \eqref{Discret_Equat}–\eqref{OtroAux4} constitutes the discrete version of the problem under investigation. The conditions ensuring convergence and stability of the adopted numerical method are well established in the literature \cite{Morton05}, and, for the specific case considered in this work, they are expressed as follows:
\begin{equation}
\begin{cases}
\label{cond_est}
\left(\dfrac{\beta_1^x \,\Delta t }{2 \, \Delta x}\right)^2 < 2 \, \dfrac{\alpha_1 \,\Delta t }{(\Delta x)^2} < 1, \qquad
\left(\dfrac{\beta_2^x \,\Delta t }{2 \, \Delta x}\right)^2 < 2 \, \dfrac{\alpha_2 \,\Delta t }{(\Delta x)^2} < 1, \\
\left(\dfrac{\beta_1^y \,\Delta t }{2 \, \Delta y}\right)^2 < 2 \, \dfrac{\alpha_1 \,\Delta t }{(\Delta y)^2} < 1, \qquad
\left(\dfrac{\beta_2^y \,\Delta t }{2 \, \Delta y}\right)^2 < 2 \, \dfrac{\alpha_2 \,\Delta t }{(\Delta y)^2} < 1.
\end{cases}
\end{equation}

Provided that these conditions are satisfied, the discretized system defined by \eqref{Discret_Equat}–\eqref{OtroAux4} achieves first-order accuracy in time and second-order accuracy in space.

\section{Numerical Example}

A non-parallel computational scheme was implemented in Matlab to perform the numerical simulations. All computations were executed within a few minutes on an Intel(R) Core(TM) i7-6700K processor running at 4 GHz. Throughout the simulations, air at atmospheric pressure is considered as the working fluid. The convective heat transfer coefficients $h_A$, $h_B$, $h_1$ and $h_2$ are estimated based on the approach presented in \cite{Umbricht20conv}, while the thermal properties of the materials are taken from \cite{Cengel07} and are summarized in the following table.
\setcounter{table}{0}
\begin{table}[h!]
\begin{center}
{\begin{tabular}{lccc} \toprule
Materials  & Symbol   & $\alpha \left(\times 10^{4}\right) \, \left[m^{2}/s\right] $  & $\kappa \, \left[W/m^{\circ}C\right] $\\ \midrule
Lead       &   $Pb$     &                      0.23673                                  &              35                      \\        
Iron       &   $Fe $    &                      0.20451                                  &              73                      \\ 
Nickel     &   $Ni $    &                      0.22663                                  &              90                      \\ 
Aluminium  &   $Al$     &                      0.84010                                  &              204                     \\ 
Copper     &   $Cu $    &                      1.12530                                  &              386                     \\ \bottomrule
\end{tabular}}  
\end{center}
\vspace{-0.5cm}
\caption{Thermal properties of different materials.}
\label{Prop_Termicas}
\end{table}

\begin{xmpl}
\label{example1}

For this example the following parameters are considered:
$x_2=1 \, m$, $x_1=0.4 \, m$, $t_\infty=10800 \, s=3 \, h$, $\bbeta_1= (0.02 \, m/s,0.02 \, m/s)$ , $\bbeta_2= \left(0.02 \, m/s, 0.02 \, m/s \,\dfrac{\alpha_2}{\alpha_1}\right)$ , $\nu_1=\nu_2=-0.0003 \,\, 1/s$, $R=0.05 \, m$ .

The initial condition is null $T_{1,0}(x,y)=T_{2,0}(x,y)=0$ and the heat generation source $s(x,t)$ is a continuous and differentiable function given by:
\begin{equation}
\label{fuente}
\begin{cases}
s_1(x,y,t)= \dfrac{100}{x_1 \,w \,t_\infty^2} \, \dfrac{^{\circ}C}{m^2 \, s} \, x \, (x_1-x)\, y (w-y) t\, (t_\infty-t), \vspace{0.2cm}\\ \, (x,y,t) \in [0,x_1] \times [0,w] \times [0,t_\infty],  \\
\\
s_2(x,y,t)=\dfrac{100}{(x_2-x_1) \, w \, t_\infty^2} \, \dfrac{^{\circ}C}{m^2 \, s} \, (x-x_1) \, (x_2-x) \, y (w-y) t\, (t_\infty-t), \vspace{0.2cm}\\ \, (x,t) \in [x_1,x_2] \times [0,w]  \times [0,t_\infty].
\end{cases}
\end{equation}
\end{xmpl}

The thermal source applied to the body plays a crucial role, as it directly determines the overall shape of the resulting temperature profiles. The function defined in equation \eqref{fuente} is especially relevant because it characterizes a heat generation process that initiates at the center of each layer and decreases gradually toward the edges, where no heat is generated. Notably, this thermal source continuously injects energy into the system, with the rate of heat generation increasing over time until it reaches a peak at \( t = 1.5 \, h \). After this point, the heat input begins to decline, eventually reaching zero at \( t = 3 \, h \).

\begin{figure}[h!]
\centering
\begin{overpic}[width=0.495\textwidth]{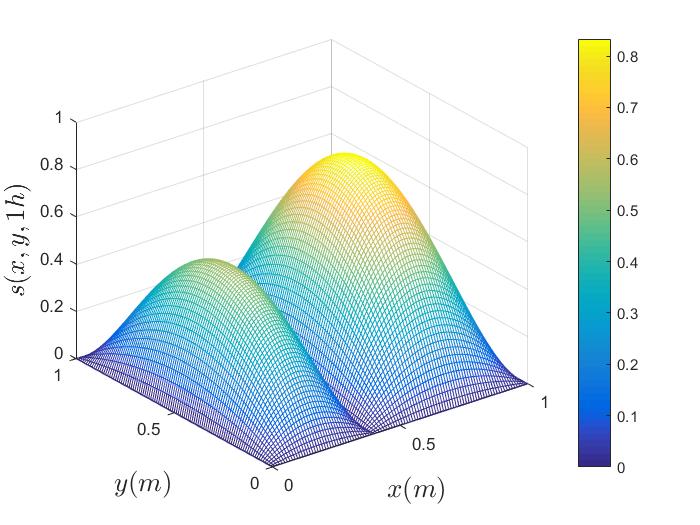}
    \put(0.8,49){\tiny \rotatebox{90}{$(^\circ C/ s)$}} 
		\put(79,72){\tiny {$(^\circ C/ s)$}} 
\end{overpic}
\begin{overpic}[width=0.495\textwidth]{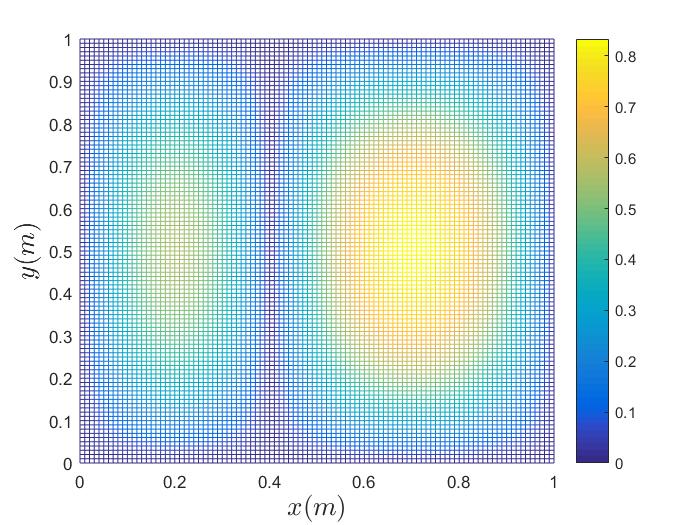}
    \put(79,72){\tiny{$(^\circ C/ s)$}} 
		\end{overpic}
\caption{Heat source at $t=1 \, h$.}
\label{Heat_source1}
\end{figure}

\begin{figure}[h!]
\centering
\begin{overpic}[width=0.495\textwidth]{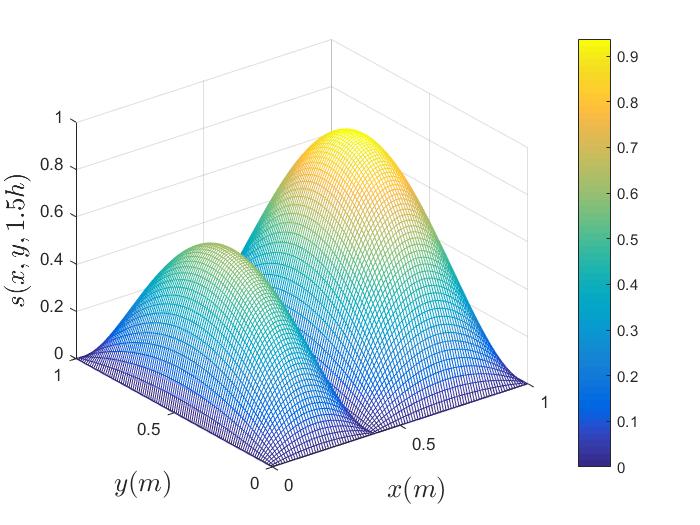}
    \put(0.8,51){\tiny \rotatebox{90}{$(^\circ C/ s)$}} 
		\put(79,72){\tiny {$(^\circ C/ s)$}} 
\end{overpic}
\begin{overpic}[width=0.495\textwidth]{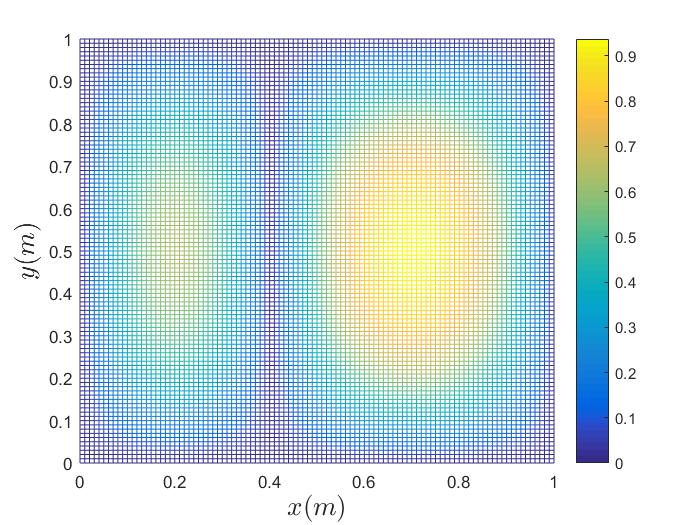}
    \put(79,72){\tiny{$(^\circ C/ s)$}} 
		\end{overpic}
\caption{Heat source at $t=1.5 \, h$.}
\label{Heat_source2}
\end{figure}

\begin{figure}[h!]
\centering
\begin{overpic}[width=0.495\textwidth]{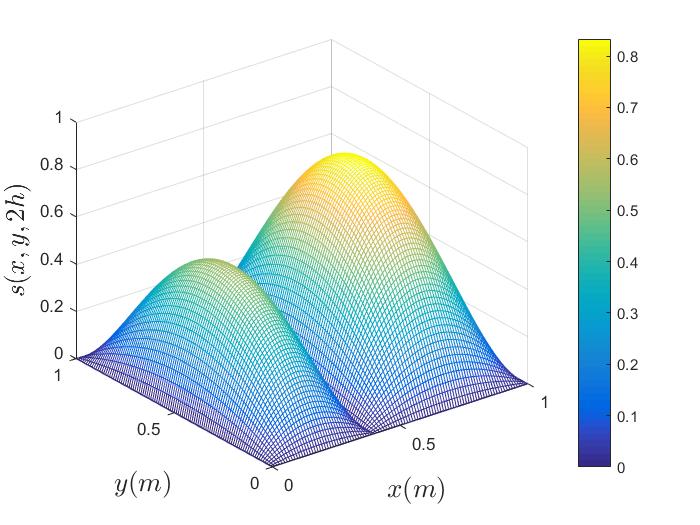}
    \put(0.8,50){\tiny \rotatebox{90}{$(^\circ C/ s)$}} 
		\put(79,72){\tiny {$(^\circ C/ s)$}} 
\end{overpic}
\begin{overpic}[width=0.495\textwidth]{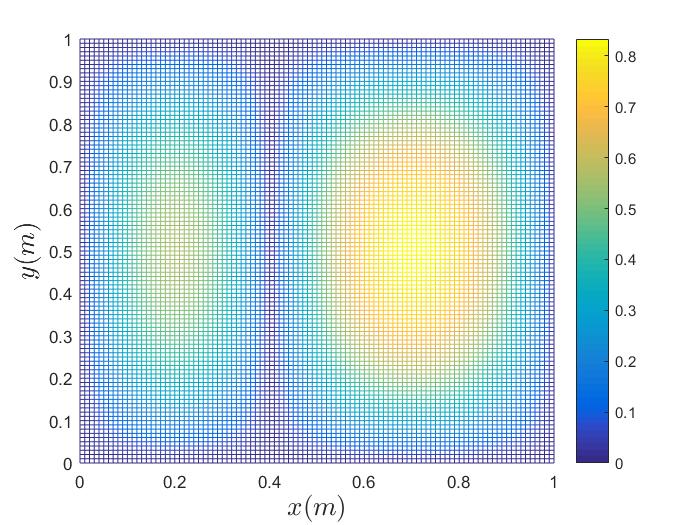}
    \put(79,72){\tiny{$(^\circ C/ s)$}} 
		\end{overpic}
\caption{Heat source at $t=2 \, h$.}
\label{Heat_source3}
\end{figure}

Figures \ref{Heat_source1}, \ref{Heat_source2}, and \ref{Heat_source3} illustrate the evolution of the source function at various time steps. In all cases, the maximum heating intensity is clearly located at the center of each layer.


\begin{figure}[h!]
\centering
\begin{overpic}[width=0.495\textwidth]{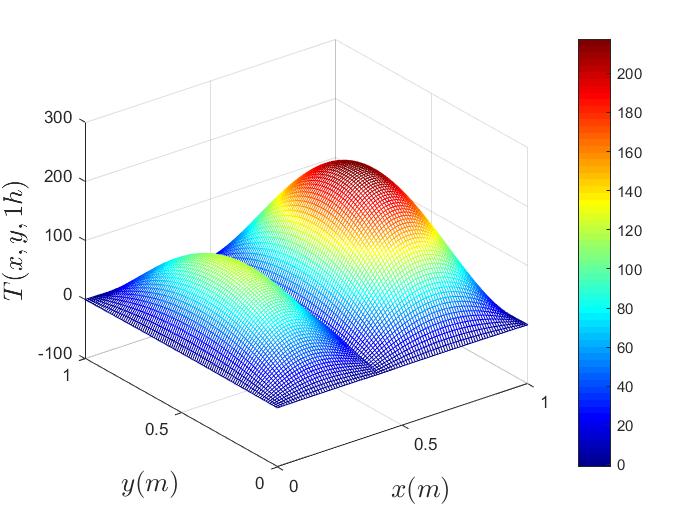}
    \put(0.5,50){\tiny \rotatebox{90}{$(^\circ C)$}} 
		\put(81,72){\tiny {$(^\circ C)$}} 
\end{overpic}
\begin{overpic}[width=0.495\textwidth]{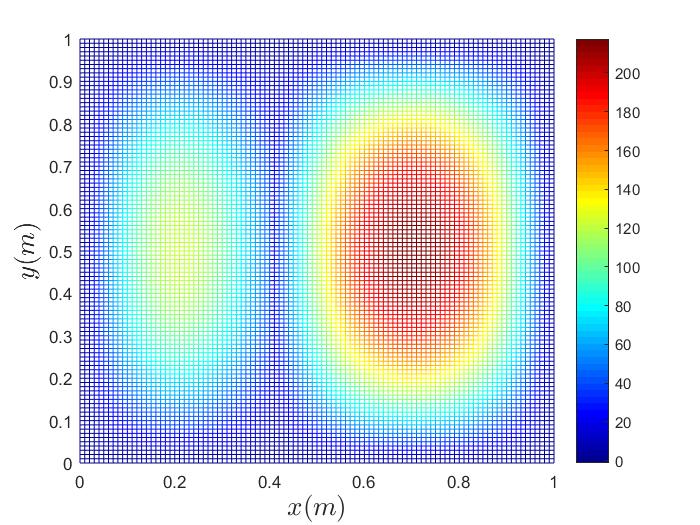}
    \put(81,72){\tiny{$(^\circ C)$}} 
		\end{overpic}
\caption{Temperature field in a $Pb-Fe$ bilayer body at $t=1 \, h$.}
\label{Temp1}
\end{figure}

\begin{figure}[h!]
\centering
\begin{overpic}[width=0.495\textwidth]{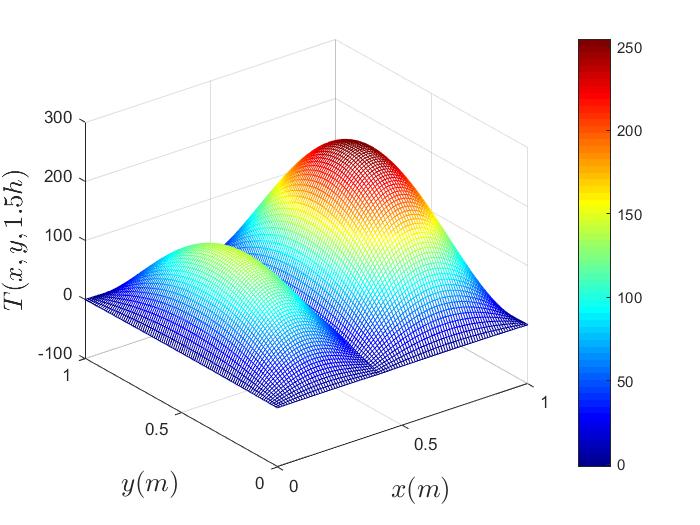}
    \put(0.5,51.5){\tiny \rotatebox{90}{$(^\circ C)$}} 
		\put(81,72){\tiny {$(^\circ C)$}} 
\end{overpic}
\begin{overpic}[width=0.495\textwidth]{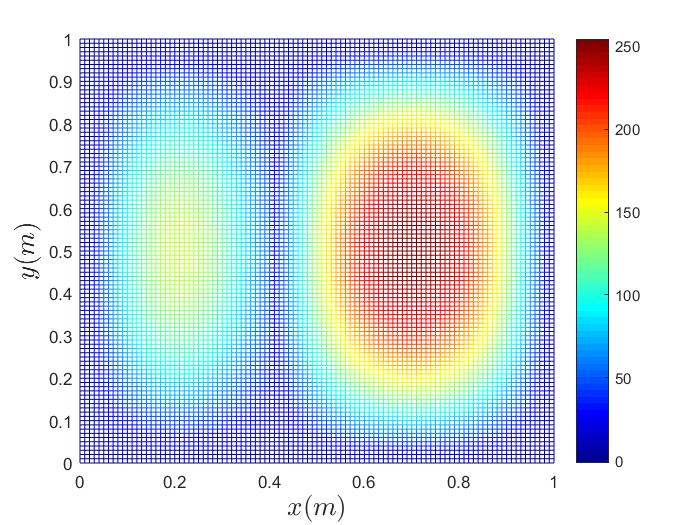}
    \put(81,72){\tiny{$(^\circ C)$}} 
		\end{overpic}
\caption{Temperature field in a $Pb-Fe$ bilayer body at $t=1.5 \, h$.}
\label{Temp2}
\end{figure}

\begin{figure}[h!]
\centering
\begin{overpic}[width=0.495\textwidth]{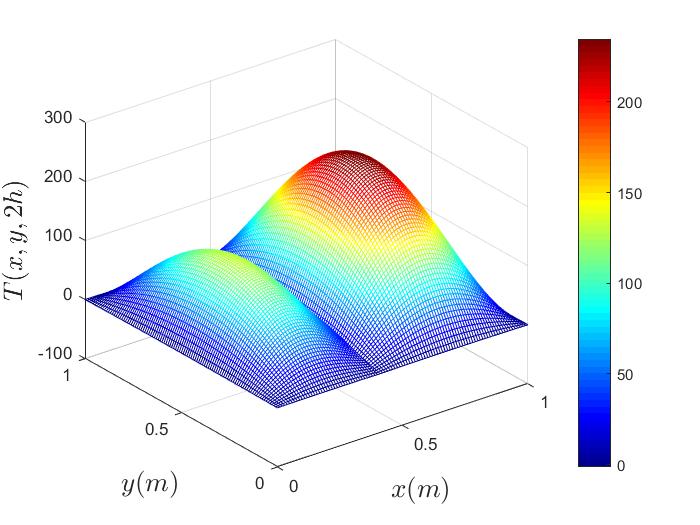}
    \put(0.5,50){\tiny \rotatebox{90}{$(^\circ C)$}} 
		\put(81,72){\tiny {$(^\circ C)$}} 
\end{overpic}
\begin{overpic}[width=0.495\textwidth]{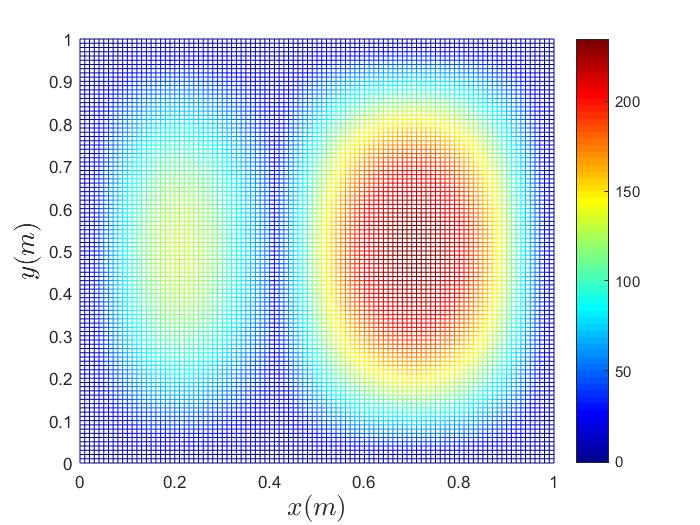}
    \put(81,72){\tiny{$(^\circ C)$}} 
		\end{overpic}
\caption{Temperature field in a $Pb-Fe$ bilayer body at $t=2 \, h$.}
\label{Temp3}
\end{figure}

Figures \ref{Temp1}, \ref{Temp2}, and \ref{Temp3} display the temperature distributions at various time instants for a bilayer  $Pb - Fe$ material. As a result of the specific form of the thermal source, the maximum temperature occurs at the center of each layer when $t = 1.5 \, h$. The temperature profiles exhibit a shape that closely mirrors that of the source term, since it is the main contributor of heat to the system.


\begin{figure}[h!]
\centering
\begin{overpic}[width=0.495\textwidth]{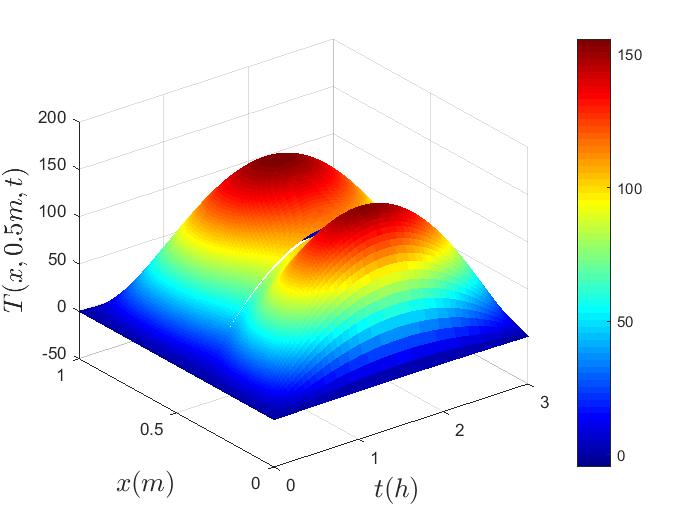}
    \put(0.5,52){\tiny \rotatebox{90}{$(^\circ C)$}} 
		\put(81,72){\tiny {$(^\circ C)$}} 
\end{overpic}
\begin{overpic}[width=0.495\textwidth]{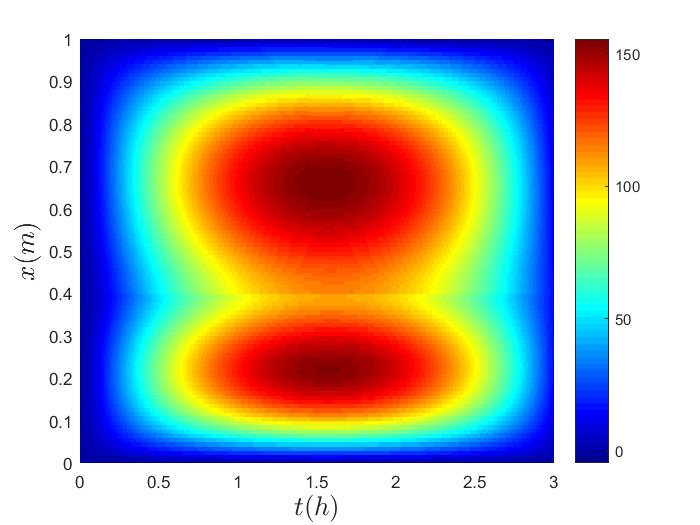}
    \put(81,72){\tiny{$(^\circ C)$}} 
		\end{overpic}
\caption{Temperature field in a $Fe-Cu$ bilayer body at $y=0.5 \, m$.}
\label{Temp4}
\end{figure}

\begin{figure}[h!]
\centering
\begin{overpic}[width=0.495\textwidth]{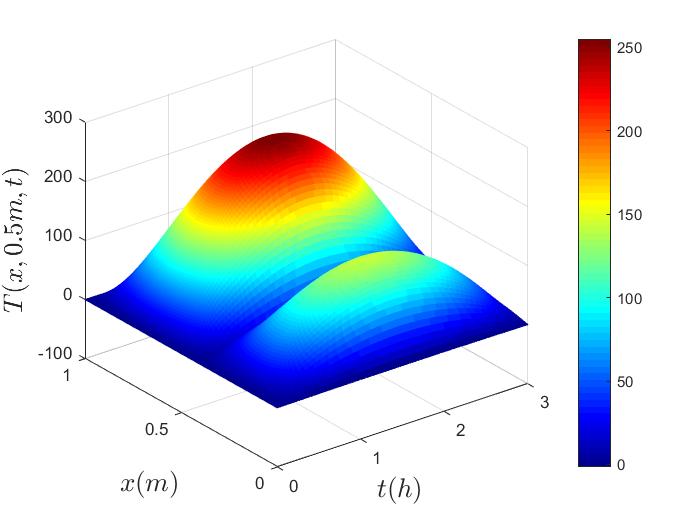}
    \put(0.5,52){\tiny \rotatebox{90}{$(^\circ C)$}} 
		\put(81,72){\tiny {$(^\circ C)$}} 
\end{overpic}
\begin{overpic}[width=0.495\textwidth]{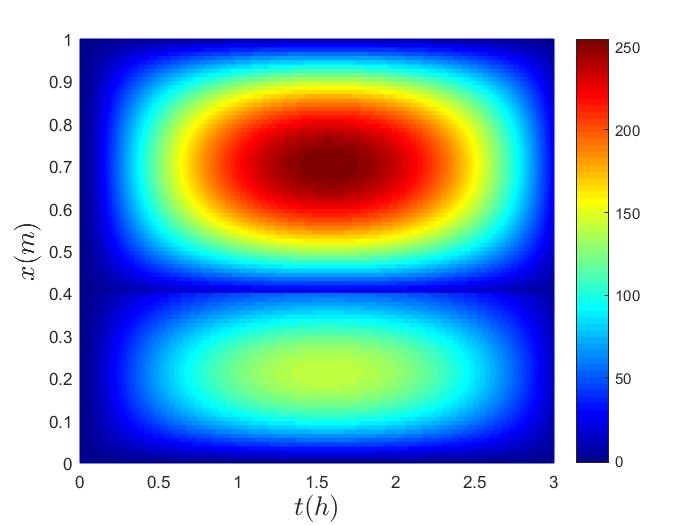}
    \put(81,72){\tiny{$(^\circ C)$}} 
		\end{overpic}
\caption{Temperature field in a $Pb-Fe$ bilayer body at $y=0.5 \, m$.}
\label{Temp5}
\end{figure}

Figures \ref{Temp4} and \ref{Temp5} present the spatiotemporal evolution of temperature at $y = 0.5 \, m$ for an $Fe-Cu$ composite and a $Pb-Fe$ composite, respectively. A temperature discontinuity is clearly visible at $x = 0.4 \, m$, becoming more prominent around $t = 1.5 \, h$. In Figure \ref{Temp4}, the more intense red hue compared to Figure \ref{Temp5} indicates higher temperatures, which is consistent with the greater thermal conductivity of $Fe$ relative to $Pb$. This behavior is in agreement with the expected physical response of the system.


\begin{figure}[h!]
\centering
\begin{overpic}[width=0.495\textwidth]{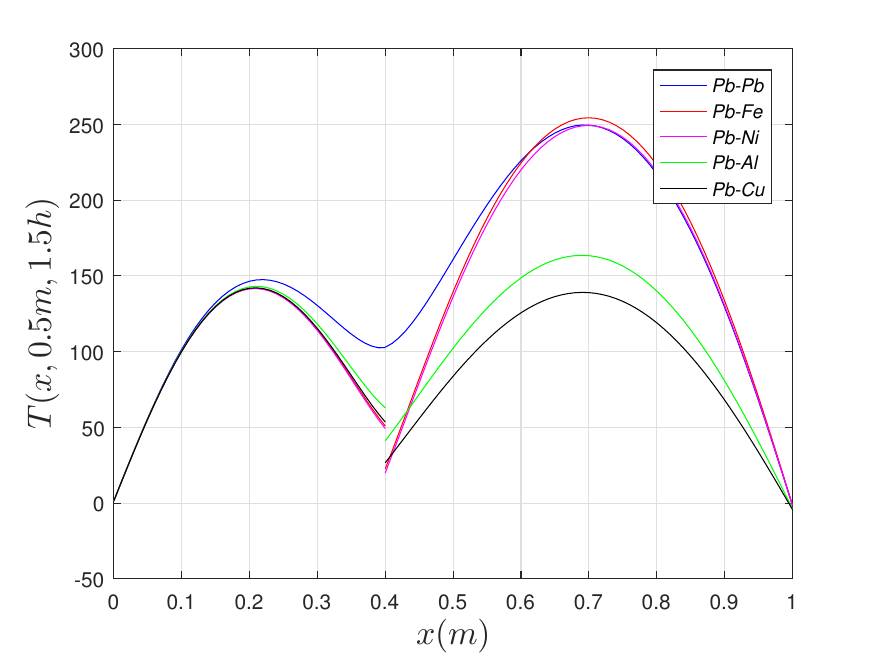}
    \put(2.8,53){\tiny \rotatebox{90}{$(^\circ C)$}} 
\end{overpic}
\begin{overpic}[width=0.495\textwidth]{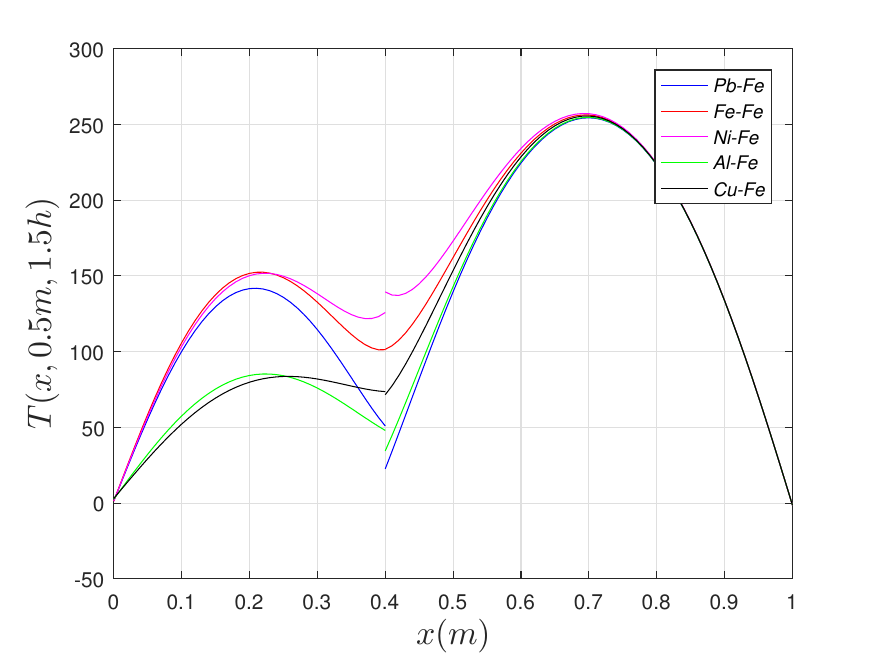}
    \put(2.8,53){\tiny \rotatebox{90}{$(^\circ C)$}} 
\end{overpic}
\caption{Temperature at $y = 0.5 \, m$ and $t = 1.5 \, h$: $Pb$-Material (left) and Material-$Fe$ (right).}
\label{TempMs}
\end{figure}

Figure \ref{TempMs} displays the spatial temperature distributions at $y = 0.5 \, m$ and $t = 1.5 \, h$. The left panel shows the results for a $Pb-Material$ configuration, while the right panel corresponds to a $Material-Fe$ arrangement. In both scenarios, a temperature discontinuity is observed at the interface. The magnitude of this discontinuity increases with the absolute difference between the ratios of thermal conductivity and thermal diffusivity of the constituent materials in the bilayer. This behavior, as shown in Table \ref{cocientes}, is consistent with the expected physical response of the system described in Eq. \eqref{Cond_Interf}.


\begin{table}[h!]
\begin{center}
{\begin{tabular}{lcc} \toprule
 Materials  & $\left|\frac{\kappa_1}{\alpha_1}-\frac{\kappa_2}{\alpha_2}\right| \left(\times 10^{-4}\right) \, \left[J/m^3 \, ^\circ C \right] $  & $\left|T_1(x_1,y,t)-T_2(x_1,y,t)\right| \, \left[^{\circ}C\right] $\\ \midrule
 $Pb-Pb$  &                 0                                  &              0.65                      \\   
$Pb-Al $ &                 94.98                                  &              21.82                    \\ 
$Pb-Cu $ &                 195.17                                  &              26.58                     \\      
 $Pb-Fe$  &                 209.10                                  &              28.34                      \\ 
$Pb-Ni$  &                  249.27                                  &              29.71                      \\ 
 $Fe-Fe$  &                 0                                  &              0.34                     \\ 
$Cu-Fe $ &                 13.93                                  &              2.35                     \\ 
 $Ni-Fe$  &                 40.17                                  &              12.49                     \\ 
 $Al-Fe $ &                 114.12                                  &              15.02                     \\ \bottomrule
 \end{tabular}}  
\end{center}
\vspace{-0.5cm}
\caption{Temperature discontinuities at the interface for $y = 0.5\,m$ and $t = 1.5\,h$.}
\label{cocientes}
\end{table}

\begin{figure}[h!]
\centering
\begin{overpic}[width=0.495\textwidth]{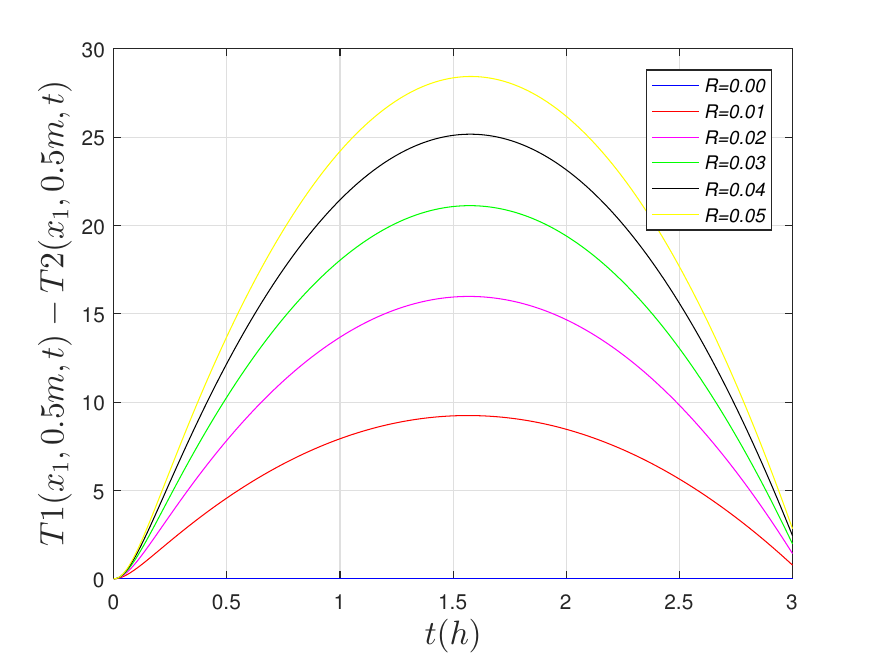}
    \put(4.4,66.5){\tiny \rotatebox{90}{$(^\circ C)$}} 
\end{overpic}
\begin{overpic}[width=0.495\textwidth]{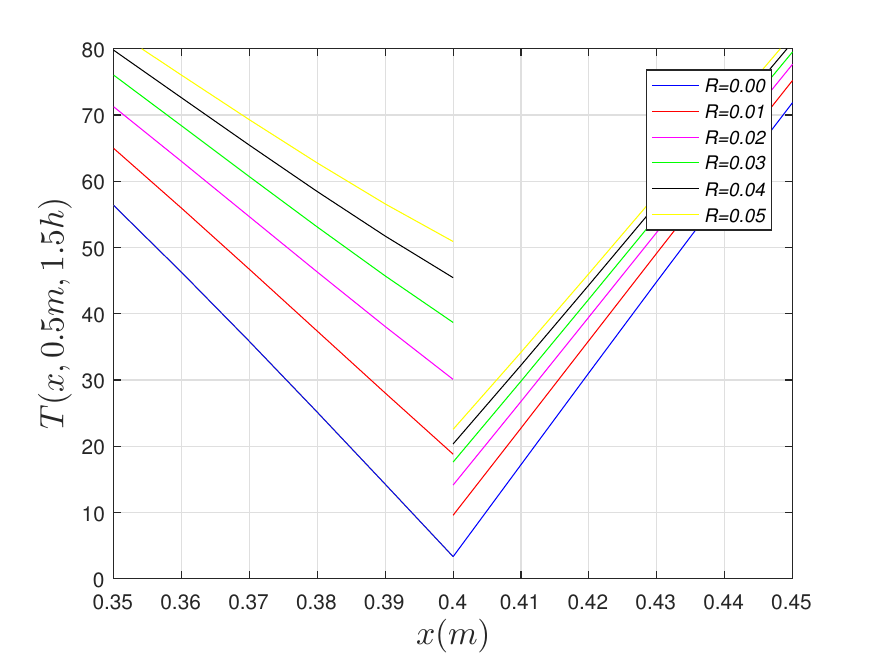}
    \put(4.2,53){\tiny \rotatebox{90}{$(^\circ C)$}} 
\end{overpic}
\caption{Temporal distribution of temperature differences at the interface at $y = 0.5 \, m$ (left). Spatial temperature distribution at $y = 0.5 \, m$ and $t = 1.5 \, h$ (right), for $Pb-Fe$, considering different thermal resistances.}
\label{TempRs}
\end{figure}

Finally, Figure \ref{TempRs} illustrates the spatial distribution of the temperature gap at the interface of a $Pb-Fe$ composite for various thermal resistance values. It is evident that this temperature gap widens as the thermal resistance increases. Additionally, the most significant differences occur around $t = 1.5 \, h$.


It is worth mentioning that the numerical results obtained in Figures \ref{TempMs}--\ref{TempRs} are consistent with those reported in \cite{Umbricht25}, where an equivalent problem is analyzed in a one-dimensional setting. 

\begin{nt}
Since a stable and convergent numerical scheme is employed with an appropriate integration step (as demonstrated in the literature), analogous numerical scheme configurations will indeed yield analogous results.
\end{nt}

\begin{nt}
The results presented in this article are applicable to any type of material, provided that the specified conditions and assumptions are upheld. This applicability is due to the fact that both the analytical and numerical solutions depend solely on the thermal conductivity and diffusivity coefficients of the materials.
\end{nt}

\section{Conclusions}

This work has provided a comprehensive mathematical and computational study of transient heat transfer in a two-dimensional, two-layer composite body. The physical model includes the effects of diffusion, advection, internal heat generation or loss proportional to temperature, and heat input from external sources. A key feature of the formulation is the inclusion of thermal resistance at the material interface, allowing for discontinuities in temperature—a phenomenon often overlooked in simpler models.

An analytical solution was obtained by applying a suitable dimensionless formulation and classical solution techniques, including separation of variables, Fourier series expansions, and superposition. The rigorously derived analytical solution encompass general cases of interface problems in bilayer materials with thermal resistance. Although their formulation is complex and difficult to manipulate, they represent a significant contribution, as they account for various possible scenarios. In particular, it is shown to reduce to previously reported results when interface resistance and source terms are neglected. Moreover, these solutions may serve as a foundation for the development of new numerical methods that incorporate such expressions. In the present manuscript, the aim is to present the analytical expressions derived for solving the problem in its general form, while the development of the aforementioned approximations is left for future research.

Beyond its theoretical relevance, the analytical formulation also provides a valuable tool for applications and optimization. Because the closed-form expressions explicitly reflect the influence of material properties, interfacial resistance, and boundary conditions, they enable systematic parametric studies without resorting to repeated simulations. This feature can assist in identifying optimal layer thicknesses, material combinations, or interface characteristics to enhance thermal performance. Such analyses are particularly relevant in engineering contexts such as thermal barrier coatings, electronic packaging, or multilayer insulation systems, where efficient control of heat transfer is critical.

In parallel, a second-order accurate finite difference method was developed for the numerical approximation of the problem. The method, based on a forward scheme in time and centered (or directional) differences in space, was adapted to treat interface discontinuities and boundary conditions effectively. Stability and convergence properties were ensured through criteria established in the numerical analysis literature.

The numerical experiments confirmed that the method captures the main physical phenomena involved: temperature accumulation near the center of each layer due to the structure of the external heat source, the impact of material properties on heat propagation, and the role of thermal resistance in creating sharp temperature gradients at the interface. The method was implemented efficiently in a non-parallel Matlab code, showing good performance for realistic parameter values.

In summary, this study offers both an analytical benchmark and a reliable numerical approach for modeling multilayer heat transfer problems with interfacial resistance and internal sources. The methodology can be extended to more complex material configurations or applied in inverse analysis tasks, such as parameter identification. Future work may consider the development of parallel computational schemes or the integration of temperature-dependent material properties for enhanced physical fidelity.

\appendix

\section{Study of the orthogonality relationship} \label{Ortogonalidad}

In this Appendix, we aim to derive the orthogonality condition that applies to this problem. This result is crucial for determining the sequence $K_{n,p}$ in \eqref{CI}. As illustrated in \eqref{fs}, for the four indices $n$, $j$, $p$, and $q$, the functions $f_{1,n,p}$, $f_{1,j,q}$, $f_{2,n,p}$, $f_{2,j,q}$, $u_{1,p}$, $u_{1,q}$, $u_{2,p}$, and $u_{2,q}$ must fulfill the following condition:
\begin{equation}
\label{f1nj}
\begin{cases} 
\bar{\alpha}  \left(f''_{1,n,p}(\bar{x}) \, u_{1,p}(\bar{y})+f_{1,n,p}(\bar{x}\right) \, u''_{1,p}(\bar{y}))+\psi_1 \,f_{1,n,p}(\bar{x}) \, u_{1,p}(\bar{y}) = - \lambda_{n,p}^2 \, f_{1,n,p}(\bar{x}) \,u_{1,p}  (\bar{y}) , \\ 
\bar{\alpha}  \left(f''_{1,j,q}(\bar{x}) \, u_{1,q}(\bar{y})+f_{1,j,q}(\bar{x}) \, u''_{1,q}(\bar{y})\right)+\psi_1 \,f_{1,j,q}(\bar{x}) \, u_{1,q}(\bar{y}) = - \lambda_{j,q}^2 \, f_{1,j,q}(\bar{x})\, u_{1,q}  (\bar{y}) ,
\end{cases}
\end{equation}
and
\begin{equation}
\label{f2nj}
\begin{cases} 
 f''_{2,n,p}(\bar{x}) \, u_{2,p}(\bar{y})+f_{2,n,p}(\bar{x}) \, u''_{2,p}(\bar{y}) + \psi_2 \,f_{2,n,p}(\bar{x}) \,u_{2,p}(\bar{y})= - \lambda_{n,p}^2 \,f_{2,n,p}(\bar{x}) \, u_{2,p}(\bar{y}),  \\ 
 f''_{2,j,q}(\bar{x}) \, u_{2,q}(\bar{y})+f_{2,j,q}(\bar{x}) \, u''_{2,q}(\bar{y}) + \psi_2 \,f_{2,j,q}(\bar{x}) \,u_{2,q}(\bar{y})= - \lambda_{j,q}^2 \,f_{2,j,q}(\bar{x}) \, u_{2,q}(\bar{y}).
\end{cases}
\end{equation}

We begin by multiplying the first equation of \eqref{f1nj} by $f_{1,j,q} , u_{1,q}$ and the second by $f_{1,n,p} , u_{1,p}$. Similarly, the first equation of \eqref{f2nj} is multiplied by $f_{2,j,q} , u_{2,q}$ and the second by $f_{2,n,p} , u_{2,p}$. For ease of notation, the functional dependencies of the functions are omitted. This results in the following:
\begin{equation}
\label{f1njbis}
\begin{cases} 
\bar{\alpha}  \left(f''_{1,n,p} \, u_{1,p}+f_{1,n,p} \, u''_{1,p}\right) f_{1,j,q} \, u_{1,q}+\psi_1 \,f_{1,n,p} \, u_{1,p} \, f_{1,j,q} \, u_{1,q} = - \lambda_{n,p}^2 \, f_{1,n,p} \,u_{1,p} \, f_{1,j,q} \, u_{1,q}  , \\ 
\bar{\alpha}  \left(f''_{1,j,q} \, u_{1,q}+f_{1,j,q} \, u''_{1,q}\right) f_{1,n,p} \, u_{1,p}+\psi_1 \,f_{1,j,q} \, u_{1,q} \, f_{1,n,p} \, u_{1,p}= - \lambda_{j,q}^2 \, f_{1,j,q}\, u_{1,q} \, f_{1,n,p} \, u_{1,p} ,
\end{cases}
\end{equation}
and
\begin{equation}
\label{f2njbis}
\begin{cases} 
 \left(f''_{2,n,p} \, u_{2,p}+f_{2,n,p} \, u''_{2,p} \right) f_{2,j,q} \, u_{2,q}+ \psi_2 \,f_{2,n,p} \,u_{2,p} \, f_{2,j,q} \, u_{2,q}= - \lambda_{n,p}^2 \,f_{2,n,p} \, u_{2,p} \, f_{2,j,q} \, u_{2,q},  \\ 
 \left(f''_{2,j,q} \, u_{2,q}+f_{2,j,q} \, u''_{2,q}\right) \, f_{2,n,p} \, u_{2,p} + \psi_2 \,f_{2,j,q} \,u_{2,q} \, f_{2,n,p} \, u_{2,p}= - \lambda_{j,q}^2 \,f_{2,j,q} \, u_{2,q} \, f_{2,n,p} \, u_{2,p}.
\end{cases}
\end{equation}

The difference between the two expressions in \eqref{f1njbis} is then taken, and similarly, the same procedure is applied to \eqref{f2njbis}.
\begin{equation}
\begin{split}
\label{fnjbisa}
&\bar{\alpha} \left( f''_{1,n,p} \, u_{1,p} \, f_{1,j,q} \, u_{1,q} + f_{1,n,p} \, u''_{1,p} \, f_{1,j,q} \, u_{1,q} - f''_{1,j,q} \, u_{1,q} \, f_{1,n,p} \, u_{1,p} - f_{1,j,q} \, u''_{1,q} \, f_{1,n,p} \, u_{1,p} \right) \\
&= \left( \lambda_{j,q}^2 - \lambda_{n,p}^2 \right) f_{1,n,p} \, u_{1,p} \, f_{1,j,q} \, u_{1,q}
\end{split}
\end{equation}
and
\begin{equation}
\begin{split}
\label{fnjbisb}
& f''_{2,n,p} \, u_{2,p} \,f_{2,j,q} \, u_{2,q}+f_{2,n,p} \, u''_{2,p} \,f_{2,j,q} \, u_{2,q}-f''_{2,j,q} \, u_{2,q} \,f_{2,n,p} \, u_{2,p}-f_{2,j,q} \, u''_{2,q} \,f_{2,n,p} \, u_{2,p} \\&= \left(\lambda_{j,q}^2- \lambda_{n,p}^2\right)  f_{2,n,p} \,u_{2,p} \, f_{2,j,q} \, u_{2,q}  ,
\end{split}
\end{equation}
the equations \eqref{fnjbisa}-\eqref{fnjbisb} are conveniently rewritten, and the first equation is multiplied by $\dfrac{\varphi \,\phi-\eta \,\mu}{\bar{\alpha}}$.
\begin{equation}
\label{fnjbis2a}
\begin{split} 
& (\varphi \,\phi-\eta \,\mu )\left\{ \,u_{1,p} \,u_{1,q}  \left[f'_{1,n,p} \,f_{1,j,q}-f_{1,n,p} \,f'_{1,j,q}\right]'+f_{1,n,p} \,f_{1,j,q} \left[u'_{1,p} \,u_{1,q}-u_{1,p} \,u'_{1,q}\right]'\right\} \\&= \dfrac{\varphi \,\phi-\eta \,\mu}{\bar{\alpha}} \left(\lambda_{j,q}^2- \lambda_{n,p}^2\right)  f_{1,n,p} \,u_{1,p} \, f_{1,j,q} \, u_{1,q}, \\ 
\end{split}
\end{equation}
\begin{equation}
\label{fnjbis2b}
\begin{split} 
& u_{2,p} \,u_{2,q}  \left[f'_{2,n,p} \,f_{2,j,q}-f_{2,n,p} \,f'_{2,j,q}\right]'+f_{2,n,p} \,f_{2,j,q} \left[u'_{2,p} \,u_{2,q}-u_{2,p} \,u'_{2,q}\right]' \\&= \left(\lambda_{j,q}^2- \lambda_{n,p}^2\right)  f_{2,n,p} \,u_{2,p} \, f_{2,j,q} \, u_{2,q}, \\ 
\end{split}
\end{equation}

the equations \eqref{fnjbis2a}-\eqref{fnjbis2b} are integrated over their respective intervals of definition and then summed. This results in:
\begin{equation}
\label{fnjbis3}
\begin{split}
&\left(\lambda_{j,q}^2 - \lambda_{n,p}^2\right) \left[\int_0^{\bar{x}_1} \int_0^{\bar{w}} \dfrac{\varphi \,\phi - \eta \,\mu}{\bar{\alpha}}  u_{1,p} \, u_{1,q}  \,f_{1,n,p} \, f_{1,j,q} \, d\bar{y} \, d\bar{x} +\int_{\bar{x}_1}^1 \int_0^{\bar{w}}  u_{2,p} \, u_{2,q}  \,f_{2,n,p} \, f_{2,j,q} \, d\bar{y} \, d\bar{x}\right] \\
&= \int_0^{\bar{x}_1} \int_0^{\bar{w}} \left(\varphi \,\phi - \eta \,\mu\right)   u_{1,p} \, u_{1,q}  \, \left[f'_{1,n,p} \,f_{1,j,q}-f_{1,n,p} \,f'_{1,j,q}\right]'   \, d\bar{y} \, d\bar{x} \\
&+ \int_0^{\bar{x}_1} \int_0^{\bar{w}} \left(\varphi \,\phi - \eta \,\mu\right)   f_{1,n,p} \, f_{1,j,q}  \, \left[u'_{1,p} \,u_{1,q}-u_{1,p} \,u'_{1,q}\right]'   \, d\bar{y} \, d\bar{x} \\
&+ \int_{\bar{x}_1}^1 \int_0^{\bar{w}} u_{2,p} \, u_{2,q}  \, \left[f'_{2,n,p} \,f_{2,j,q}-f_{2,n,p} \,f'_{2,j,q}\right]'   \, d\bar{y} \, d\bar{x} \\
&+ \int_{\bar{x}_1}^1 \int_0^{\bar{w}}  f_{2,n,p} \, f_{2,j,q}  \, \left[u'_{2,p} \,u_{2,q}-u_{2,p} \,u'_{2,q}\right]'   \, d\bar{y} \, d\bar{x}. 
\end{split}
\end{equation}

Then, using $u_2(\bar{y})=q \, u_1(\bar{y}) $, $\forall \bar{y} \in (0,\bar{w})$, the relation given in \eqref{par} and the boundary conditions from \eqref{fs}, one applies this to the right-hand side of equation \eqref{fnjbis3} and obtains:
\begin{equation}
\label{fnjbis4}
\begin{split}
&\left(\lambda_{j,q}^2 - \lambda_{n,p}^2\right) \left[\int_0^{\bar{x}_1} \int_0^{\bar{w}} \dfrac{\varphi \,\phi - \eta \,\mu}{\bar{\alpha}}  u_{1,p} \, u_{1,q}  \,f_{1,n,p} \, f_{1,j,q} \, d\bar{y} \, d\bar{x} +\int_{\bar{x}_1}^1 \int_0^{\bar{w}}  u_{2,p} \, u_{2,q}  \,f_{2,n,p} \, f_{2,j,q} \, d\bar{y} \, d\bar{x}\right] \\
&= \left(\bar{\varphi} \,\bar{\phi} - \bar{\eta} \,\bar{\mu}\right)  \, \left[f'_{1,n,p} \,f_{1,j,q}-f_{1,n,p} \,f'_{1,j,q}\right] \bigg|_{\bar{x}_1} \int_0^{\bar{w}} u_{2,p} \, u_{2,q} \, d\bar{y}   \\
&-  \left[f'_{2,n,p} \,f_{2,j,q}-f_{2,n,p} \,f'_{2,j,q}\right] \bigg|_{\bar{x}_1} \int_0^{\bar{w}} u_{2,p} \, u_{2,q} \, d\bar{y}   . 
\end{split}
\end{equation}
From the interface conditions in \eqref{fs} and by performing algebraic operations, the orthogonality condition is obtained. For $n \neq j$ and $p \neq q$ it is satisfied that:
\begin{equation}
\label{CondOrtogonalidad}
\left(\lambda_{j,q}^2 - \lambda_{n,p}^2\right) \left[\int_0^{\bar{x}_1} \int_0^{\bar{w}} \dfrac{\varphi \,\phi - \eta \,\mu}{\bar{\alpha}} f_{1,n,p} u_{1,p} \,f_{1,j,q} u_{1,q}   \, d\bar{y} \, d\bar{x} +\int_{\bar{x}_1}^1 \int_0^{\bar{w}} f_{2,n,p} u_{2,p} \,f_{2,j,q} u_{2,q}   \, d\bar{y} \, d\bar{x}\right] =0.
\end{equation}

\section*{Acknowledgments}

The authors would like to express their gratitude to three anonymous referees
for their constructive comments, which greatly improved the readability
of the manuscript.

This work has beem partially sponsored by the proyects 006-25CI2002
and 006-24CI1904 from Universidad Austral, Rosario, Argentina.




\end{document}